# Title:
# Triphenylamine-based interlayer with carboxyl anchoring group for tuning of charge collection interface in stabilized p-i-n perovskite solar cells and modules

## Authors:

**P.K. Sukhorukova[1,2,§], E.A. Ilicheva[1,§], P.A. Gostishchev[1], L.O. Luchnikov[1], M.M. Tepliakova[3], D.O. Balakirev[2], I.V. Dyadishchev[2], A.A. Vasilev[4], D.S. Muratov[1], Yu. N. Luponosov[2], A. Di Carlo[5] and D.S. Saranin[1]**

[1]LASE – Laboratory of Advanced Solar Energy, National University of Science and Technology "MISiS", Leninsky prospect 4, 119049 Moscow, Russia

[2] Enikolopov Institute of Synthetic Polymeric Materials of the Russian Academy of Sciences (ISPM RAS), Profsoyuznaya St. 70, Moscow, 117393, Russia

[3]Center for Energy Science and Technology, Skolkovo Institute of Science and Technology, Bolshoi blvd., 30, b. 1, Moscow, 121205, Russian Federation

[4]Department of Semiconductor Electronics and Semiconductor Physics, National University of Science & Technology MISIS, 4 Leninsky Ave., Moscow, 119049, Russia

[5]C.H.O.S.E. (Centre for Hybrid and Organic Solar Energy), Department of Electronic Engineering, University of Rome Tor Vergata, via del Politecnico 1, 00133 Rome, Italy

§: The authors contributed equally to this work

**Corresponding authors:**

Dr. Yu.N. Luponosov luponosov@ispm.ru,

Dr. Danila S. Saranin saranin.ds@misis.ru,

Prof. Aldo Di Carlo aldo.dicarlo@uniroma2.it

## Abstract:

Interface engineering is one of the most critical directions in the development of photovoltaics (PVs) based on halide perovskites. A novel triphenylamine-based hole transport material (HTM) with a carboxyl anchoring group (TPATC) was developed for tuning the interface between nanocrystalline NiO and double cation $CsCH_3(NH_2)_2PbI_{3-x}Cl_x$ absorber in p-i-n device architectures. We present a unique comprehensive confinement study of PSCs with TPATC as the self-assembled HTM, including analysis of numerical defect parameters, phase composition evolution, and up-scaling capabilities. Our investigation shows that the ultrathin TPATC interlayer effectively passivates traps, increases work-function of HTL by about ~0.2 eV, and enhances the charge carrier extraction efficiency. Advanced transient spectroscopy measurements revealed that modification of the NiO surface with TPATC in perovskite solar cells (PSCs) reduces the concentration of ionic defects by an order of magnitude. It was found that the reference devices with NiO HTL contain formamidine vacancies ($V_{FA}$) with an activation energy ($E_a$) of 0.509 eV, while for NiO/TPATC PSCs energy levels associated with I-Pb anti-sites ($E_a$= 0.766 eV) were observed. Interface engineering with TPATC allowed to reach power conversion efficiency of 20.58% for small area devices (0.15 cm$^2$) under standard AM 1.5 G conditions. Using TPATC interlayer also provided stabilized performance of PSCs under operation conditions and improved sustainability of the perovskite absorber to decomposition. After continuous light-soaking (1000 h, ISOS-L-2 protocol), NiO/TPATC devices showed a slight decrease of 2% in maximum power. In contrast, NiO PSCs demonstrated decrease in power output (>20%) after 400 h. We explored the potential of TPATC to modify interfaces in large-area perovskite solar modules (PSM, active area-64.8 cm2, 12 sub-cells). By applying slot-die-coated TPATC, the PCE at AM 1.5 G conditions increased from 13.22% for NiO PSM to 15.64% for NiO/TPATC ones. This study provides new insights into the interface engineering for p-i-n perovskite solar cells, behavior of the ionic defects and their contribution to the long-term stability.

1. **Introduction**

Halide perovskites (HP) offer strong optical absorption ($10^{-5}$ cm$^{-1}$)[1] in the visible part of the light spectra, long carrier diffusion lengths (up to several micrometers)[2], and relatively high charge-carrier mobilities ($10^{-2}$-$10^{1}$ cm$^2$ V$^{-1}$s$^{-1}$)[3], [4]. The chemical composition of halide perovskites (HPs) is represented by the general formula $ABX_3$[5], where A- stands for an organic cations $CH_3NH_3$(MA), $CH_3(NH_2)_2$(FA), or Cs; B- represents a metal cation, typically Pb; and X- denotes a halide anion I, Br, or Cl. The highest power conversion efficiency (PCE) of perovskite solar cells (PSCs) currently stands at 26.1%[6], making them competitive with the industry standards of crystalline Si cells in the photovoltaic field.

The possibility of large-scale fabrication using low-temperature solution processing techniques such as slot die coating[7] and doctor blade[8] has the potential to reduce capital expenditures (CAPEX) for mass production[9]. However, the rapid degradation of HP-based devices limits their practical use. Several intrinsic properties of HP promote material degradation under exploitation conditions (heat, light exposure, and moisture)[10]–[12]. Perovskite solar cells (PSCs) are thin-film heterostructures consisting of a nano-/micro-crystalline perovskite absorber sandwiched between n- and p-type charge transporting layers [13].

The chemical stability of interfaces in multilayer device structures play a critical role in the long-term stability of PSCs[14]–[17]. HP absorbers undergo decomposition processes upon exposure to heat and light, leading to the formation of volatile compounds (FA or MA), iodine and other byproducts[18]–[20]. This can induce intense corrosion and oxidation processes, which further affect the stability of PSCs. Antisites (I-Pb)[21], iodine interstitials[22], and metallic (lead) clusters[23] can cause deep defect states. The accumulation of defects at the interfaces leads to energy losses due to non-radiative recombination[24] and initiates electrochemical interactions within the device's structure. To mitigate defect activity at the interfaces of perovskite solar cells, the development of passivation approaches is crucial[25]. Interface engineering is an important strategy to improve the performance of perovskite solar cells (PSCs), and organic self-assembling monolayer (SAM) materials have recently been shown to be effective for this purpose. SAMs are ordered molecular structures with one or several molecules thick formed during the adsorption of active surfactants on a solid surface[26]–[28]. It can be formed on semiconductor and dielectric surfaces and has been used in a variety of technological applications, including organic electronic devices such as organic light-emitting diodes [26], organic photovoltaics[29], organic thin film transistors[30], and nonvolatile memories[31].

According to the reports[32], [33], SAMs can be efficiently utilized as hole-transport materials (HTMs) for PSCs, which are particularly suitable for the inverted device structures, where hole selective contact is placed at the light incident side. Hole-transport layers (HTLs) in PSCs allows to simplify and improve the hole transport from the photoactive layer to the anode and to create an energy barrier that prevents electron transfer and resulted charge recombination. The SAM formation as HTL is associated with the occurrence of a specific interaction between reactive groups (so-called anchoring groups) of the functional molecule and, for example, the substrate material. In general, the molecules used for SAM materials consist of three main structural parts: (i) the anchoring group, which is linked through (ii) a spacer group to the (iii) terminal unit, which is responsible for the charges transport and energy levels. Frequently used anchoring groups are residues of organic and organophosphorus acids, since they form quite strong bonds with metal oxides. Derivatives of triphenylamine (TPA) and carbazole are often used as a terminal group in SAM materials because of their suitable HOMO energy levels for hole transport and good charge-transfer properties[34]–[36].

Single-junction PSCs and perovskite-based tandem solar cells employing SAM as HTLs have achieved power conversion efficiencies (PCEs) over 24%[32], [37] and 29%, respectively. SAMs can be fine-tuned by varying the size of molecules and can be optimized by modulating the energy level, wettability, and surface interaction[28], [38]. Meanwhile, the role of SAMs materials in stabilizing interfaces and changes of the defect parameters in perovskite solar cells has not been thoroughly explored. The application of SAM materials is promising for engineering interfaces between perovskite absorbers and inorganic charge-transport layers. Thin-films of nickel oxide are standard materials for p-i-n perovskite solar cells. NiO is a wide-area material ($E_g$>3.5 eV)[39], has a large value of ionization potential (>5.0 eV)[40]–[42], and relatively high hole mobility for nanocrystalline layers[43]. However, the presence of surface states and charged defects does not provide passivation of the interface with perovskite absorbers[44]–[48], leading to electrochemical interactions.

In this paper, we present a comprehensive study on the utilization of novel SAM material to enhance the stability and performance of a NiO-based perovskite solar cell. As a SAM material we synthesized 5-[4-(diphenylamino)phenyl]thiophene-2-carboxylic acid (**TPATC**), which is based on electron donor TPA unit, which is linked through the thiophene spacer to the carboxyl fragment group as the anchoring group. The **TPATC** can be considered as one of the simplest in terms of the synthesis SAM material. According to the literature data[49]–[51], materials of this type are known for their high stability and possibility to form sufficiently strong bonds with the substrate.

We found that the formation of the **TPATC** interlayer on the NiO surface has a substantial impact on the properties of the hole-transport stack and impacts the stability of the device performance under exploitation conditions. The use of the new SAM molecule enabled to reach the 20.8% of PCE on small-area perovskite solar cells. By engineering the NiO/perovskite interface with SAM, photocarrier lifetime has been extended while non-radiative recombination processes have been suppressed. Special attention was devoted to deep level transient spectroscopy (DLTS) performed on modified p-i-n PSCs. By analyzing the variations in defect parameters, including activation and concentration energies, we have identified that SAM engineering effectively suppresses the mobility of ionic defects and reduces the traps concentration. Ambient stability studies under light-soaking conditions revealed that, after 1000 hours, encapsulated PSCs with SAM exhibited losses of less than 5%, whereas the reference cells reached 20% losses after 560 hours. The evolution of the X-ray diffraction spectra was analyzed for the devices that were exposed to continuous illumination and heating. The samples with a SAM layer exhibited a reduction in the formation dynamics of $PbI_2$ and $CsPbI_3$ phases. Finally, we successfully employed TPATC for up-scaled solar modules (PSMs) with an active area of 65 $cm^2$ (12 sub-cells) using slot-die coating. The PCE of the champion up-scaled exceeded PCE>15% and demonstrated significant improvements compared to the reference with un-modified NiO surface.  We found that TPATC significantly increases the open-circuit voltage of perovskite solar modules and reduces contact resistance losses. Measurements under ambient low-light conditions showed that TPATC application is beneficial for suppressing trapping effects and output power at illuminances from $10^4$ to ultra-low $10^1$ Lux.

# Results and discussions

## 2.1 Synthesis

The scheme for the synthesis of **TPATC** is shown in **Fig.1a** and consists of two steps. First, the Suzuki cross-coupling reaction was carried out to prepare diphenyl[4-(2-thienyl)phenyl]amine (**3**) from (4-bromophenyl)diphenylamine (**1**) and 4,4,5,5-tetramethyl-2-(2-thienyl)-1,3,2-dioxaborolane (**2**). The introduction of carboxyl groups into the thiophene unit can be carried via oxidation of the corresponding aldehyde with silver (I) oxide in the presence of sodium hydroxide[52] However, this requires an additional reaction step to prepare the aldehyde. Therefore, we elaborated simpler way to obtain the target compound via direct carboxylation of the lithium derivative of compound **3**. Purity and the chemical structure of the obtained compounds were proven by the complex of physico-chemical methods of analysis (see **fig.S1 – S5** in Electronic supplementary information (ESI)). TPATC is a crystalline material with the high melting temperature (240 °C) and high thermal and thermal-oxidative stability (above 260 °C) (ESI, **Fig. S7** and **S8**). The optical and electrochemical properties of the compound in solution and in thin film were studied by UV-vis absorption spectroscopy and cyclic voltammetry (CV) (ESI, **fig. S9** and **S10**). TPATC absorbs light in the range of ca. 350-450 nm (**Fig.1b**) and therefore does not interfere with the absorption of light by the active layer. According to CV data, HOMO energy level of TPATC (-5.5 eV) is suitable for hole transport from the perovskite to $NiO_x$ (**Fig. 1c**).

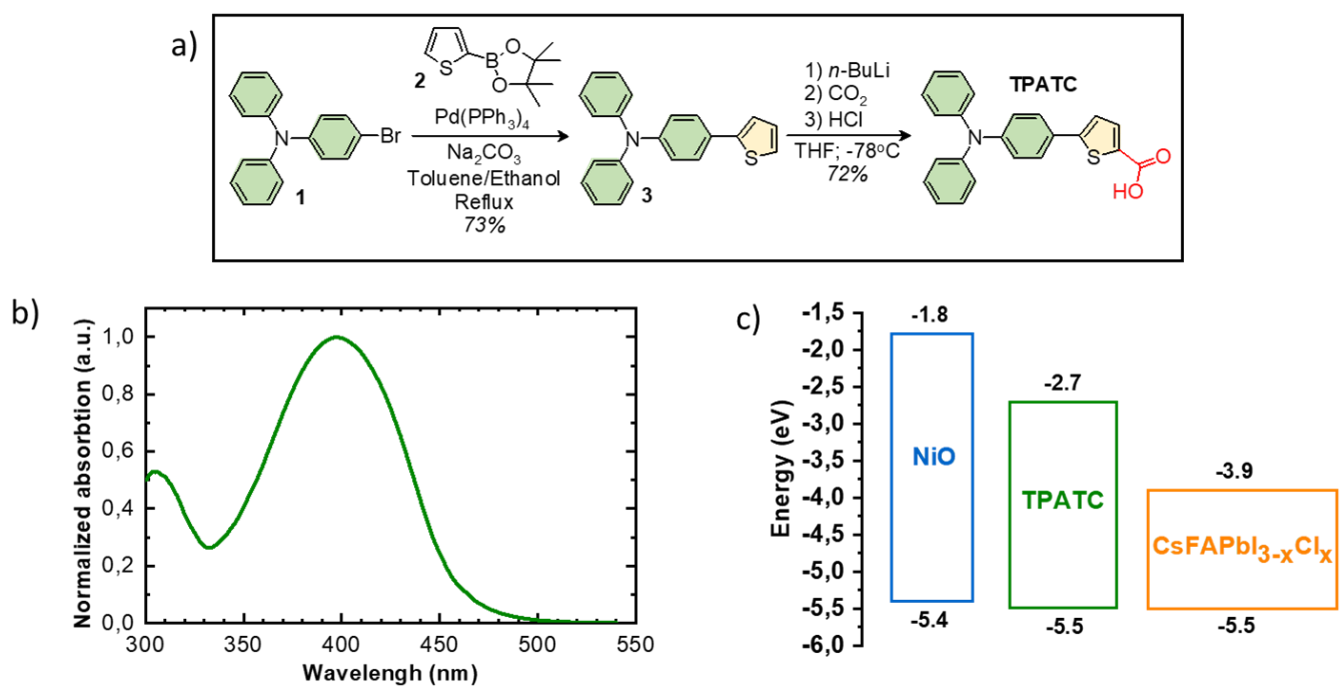

Figure 1 - Synthetic route (a), normalized UV-vis absorption spectra in thin film (b) of the **TPATC** and energy level diagram for **TPATC** compared to perovskite and NiO , where energy values of NiO and $CsFAPbI_{3-x}Cl_x$ were taken from the literature[53]–[55] (c)

To determine the specific properties of the synthesized self-assembled monolayer (SAM), the study was performed sequentially on NiO/TPATC thin-film structures, multilayer stacks with perovskite absorber, and inverted planar solar cells.

We used X-ray photoelectron spectroscopy (XPS) to investigate the surface properties of thin-film stacks - ITO/NiO and ITO/NiO/TPATC. According to the Ni2p spectra of NiO films deposited from a nickel chloride solution (**Fig. S12** in ESI), one can detect the prevalence of the multiplet characteristic of nickel oxide [56], [57]. The shape of the multiplet in the ITO/NiO and ITO/NiO/TPATC samples differs from pure NiO by an increased contribution to the spectrum of the components at BE = 855.6 eV and 861.0 eV. Deviations from the spectrum of pure NiO in the indicated binding energies are often interpreted as the contribution of the bonding of nickel with hydroxo groups formed by interaction with the atmosphere or solvent during solution-processing [58], [59] or with bonding of $Ni^{3+}$ with oxygen[60]. **Fig.2a** displays the high-resolution spectra of C1s for both sample configurations. In the ITO/NiO sample, the presence of C1s core peaks with C-O, C-C, C-H, and C=O bonds can be attributed to the absorbed organics impurities from the atmosphere [61], [62]. In the ITO/NiO/TPATC sample, the signals of C-H and C-C bonds demonstrate a significant increase in intensity. The XPS spectra analysis revealed that the atomic concentration of carbon raises from 19.5% to 48.6% with the presence of SAM coating on the surface of NiO film as shown in **Table 1**. The ITO/NiO/TPATC sample exhibits the presence of C-N and C-S bonds. The high-resolution spectra of sulfur and nitrogen, shown in **fig.2(b)** and **(c)** respectively, reveal the specific binding energies S2p at 163.8 eV and N1s at 399.2 eV. This confirms the nature of the bonding of these elements to carbon and indicates the absence of bonding to other components at the NiO surface. Both sample configurations demonstrated chlorine bound to nickel (BE = 198.3 eV), as indicated in **fig.S13** (ESI). We assume that this feature originated from the nickel chloride precursor used for the fabrication of NiO films (see experimental for the details). The surface morphology of the NiO film with SAM was analyzed using atomic force microscopy (AFM). The presence of a thin TPATC layer didn't demonstrate distinguishable changes on the nanocrystalline NiO surface, as shown in **Fig. 2(d)** and **2(e)**. The values of the mean roughness (Ra) of the NiO surface were almost identical. $R_a$ for ITO/NiO sample was 1.97 ± 0.17 nm and for ITO/NiO/TPATC was 2.0 ± 0.25 nm. Kelvin probe force microscopy (KPFM) measurements were utilized to determine the work function ($W_f$) distributions. The mean work function of the NiO/TPATC sample increased to 5.08 ± 0.1 eV from 4.86 ± 0.1eV for the reference NiO film. Photo emission spectra near the Fermi level of NiO and NiO/TPATC films is shown in **fig. S6**. Energy differences Δζ between Ef and VBM were determined as linear approximations of VBM edge. For NiO sample Δζ = 0.552 eV and 0.329 eV for NiO/TPATC. Nickel oxide band gap was found as 3.60 eV from optical absorbance spectra using Tauc plot [63] in **fig. S14** (ESI). **Fig. 2(h)** summarizes energy diagrams for NiO and NiO/TPATC HTL stacks (valence band spectra presented in **fig.S15**, ESI).

After conducting a study on hole transport layer films with TPATC, we analyzed the changes in the structural and optical properties of multilayer structures with a perovskite absorber. We used double-cation $CsCH_3(NH_2)_2PbI_{3-x}Cl_x$ ($CsFAPbI_{3-x}Cl_x$) fabricated with solution processing according to our previous work [64]. The phase composition of samples with $CsFAPbI_{3-x}Cl_x$ was investigated using X-ray diffraction (XRD) spectroscopy. **Fig. 3(a)** shows the diffractograms of perovskite films crystallized on the ITO/NiO and ITO/NiO/TPATC substrates. Both patterns show a close match with the trigonal α-FAPbI3 phase, with the space group P3m1 [65], [66] and the first peak at 20.78° corresponding to (101) crystalline plane. Peaks at 45 and 53° were attributed to the ITO substrate. Close inspection of the XRD data didn't reveal any significant differences between the samples with and without TPATC. Therefore, we conclude that modifying the NiO surface with TPATC does not affect the crystalline structure of the photoactive perovskite. Scanning electron microscopy images of the films (**Fig.S16**, ESI) reveal a close grain size distribution on the $CsFAPbI_{3-x}Cl_x$ surface with average values of 135 nm and 140 nm for NiO and NiO/TPATC samples, respectively.

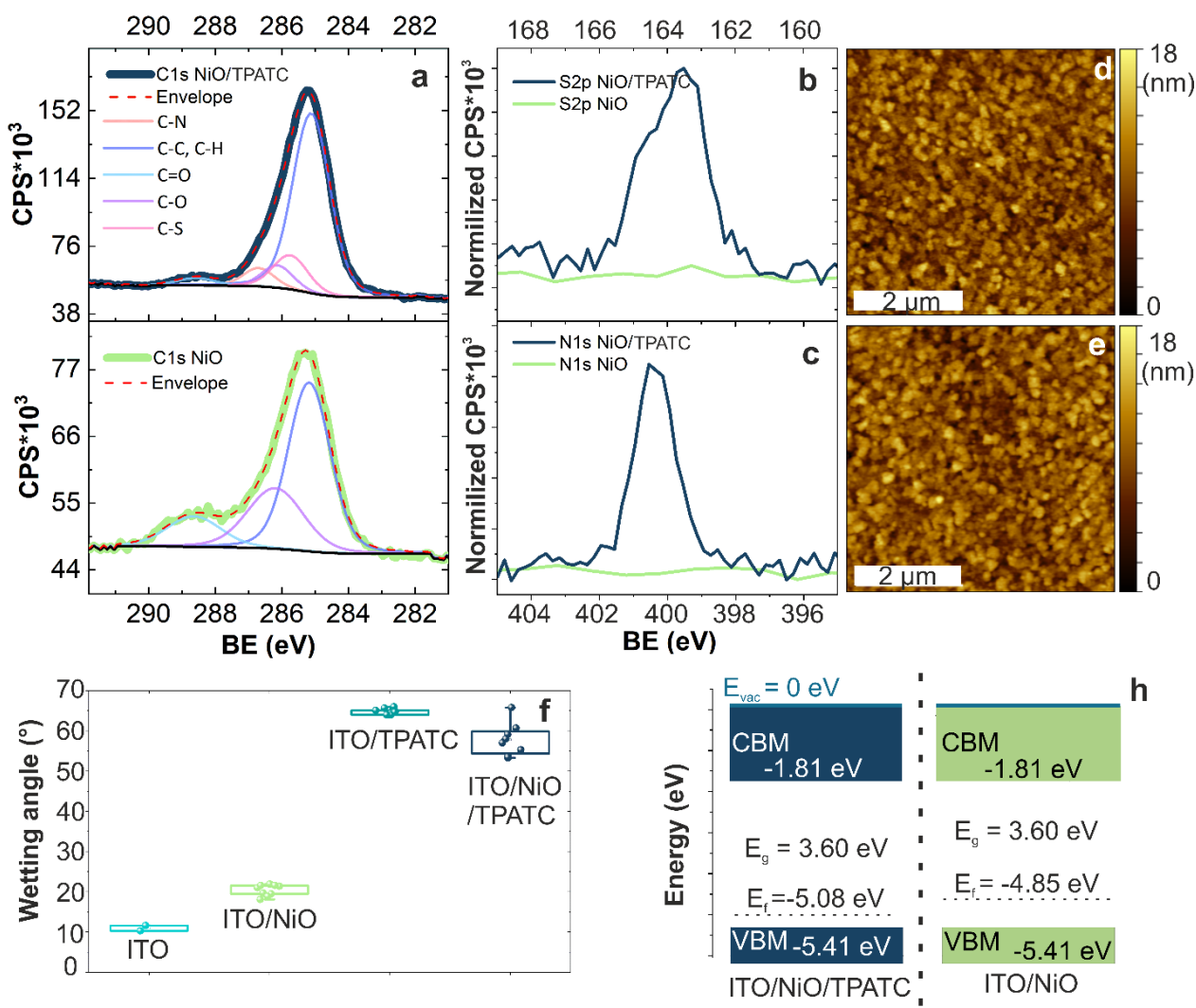

Figure 2 - characterization of ITO/NiO and ITO/NiO/TPATC stack on ITO substrate; a) XPS C1s core level spectra; b) S2p Core level; c) N1s core level; d) ITO/NiO surface morphology; e) ITO/NiO/TPATC surface morphology; f) water contact angle on films surfaces; h) band diagram of ITO/NiO and ITO/NiO/TPATC

Table 1 - atomic concentration (%) in NiO and NiO/TPATC films

|  | C | O | Ni | Cl | S | N |
|---|---|---|---|---|---|---|
| NiO | 19.45 | 44.6 | 34.09 | 1.86 | 0 | 0 |
| NiO/TPATC | 48.6 | 27.37 | 19.01 | 1.6 | 1.78 | 1.64 |

To estimate the impact of using TPATC SAM for the hole transport interface with NiO on the optical properties, we measured absorption and photoluminescence (PL) spectra (**Fig. 3(b)(c)**). The bandgap values of the multilayer stacks with $CsFAPbI_{3-x}Cl_x$ were 1.584 eV with NiO and 1.586 eV with NiO/TPATC, as extracted from a Tauc plot[63]. The PL measurements showed a single peak at 783 nm for both sample configurations. The intensity of the PL signal was approximately 40% higher for the $CsFAPbI_{3-x}Cl_x$ crystallized on the NiO/TPATC compared to the NiO reference. Given the almost identical morphology of $CsFAPbI_{3-x}Cl_x$ on NiO and NiO/TPATC, we assume that the increase in PL intensity originated from the suppression of non-radiative recombination processes at the HTL/perovskite interface with the use of SAM [67]–[72].

The carrier recombination dynamics were further characterized through time-resolved PL measurements (TRPL, **fig.3(d)**). To extract carrier lifetime PL decay curves were fit with double exponential function $I(t)=A_1 exp(-x/\tau_1)+A_2 exp(-x/\tau_2)$. Two components of this function reflect fast and slow intensity decay processes, which are associated with non-radiative trap-assisted recombination and radiative bimolecular recombination, respectively, and $A_i$ goes for decay amplitude. The average charge carrier lifetime is further calculated as $\sum A_i \tau_i^2/\sum A_i \tau_i$. To gather the reliable statistics total of five measurements were made for each sample configuration, the data is summarized in Table 2.

The median PL lifetime for the sample with configuration NiO/TPAPC/perovskite (71.3 ns) was found to be slightly lower than that for the reference NiO/perovskite (79.6 ns) suggesting more efficient charge extraction for TPAPC-containing sample.

Table 2 - TRPL data for perovskite films fabricated on NiO; NiO/TPATC

| Sample configuration | | $A_1$ | $\tau_1$ (ns) | $A_2$ | $\tau_2$ (ns) | $\tau_{average}$ (ns) |
|---|---|---|---|---|---|---|
| NiO/TPATC/ perovskite | median | 166.3 | 8.1 | 398.3 | 74.2 | 71.3 |
| | average±std | 168.4±27.5 | 8.5±1.2 | 373.1±30.5 | 78.7±12.2 | 75.42±12.5 |
| NiO/ perovskite | median | 128.3 | 6.6 | 436.4 | 81.3 | 79.6 |
| | average±std | 142.7±30. | 6.2±0.8 | 415.4±18.8 | 80.9±8.4 | 79.0±8.0 |

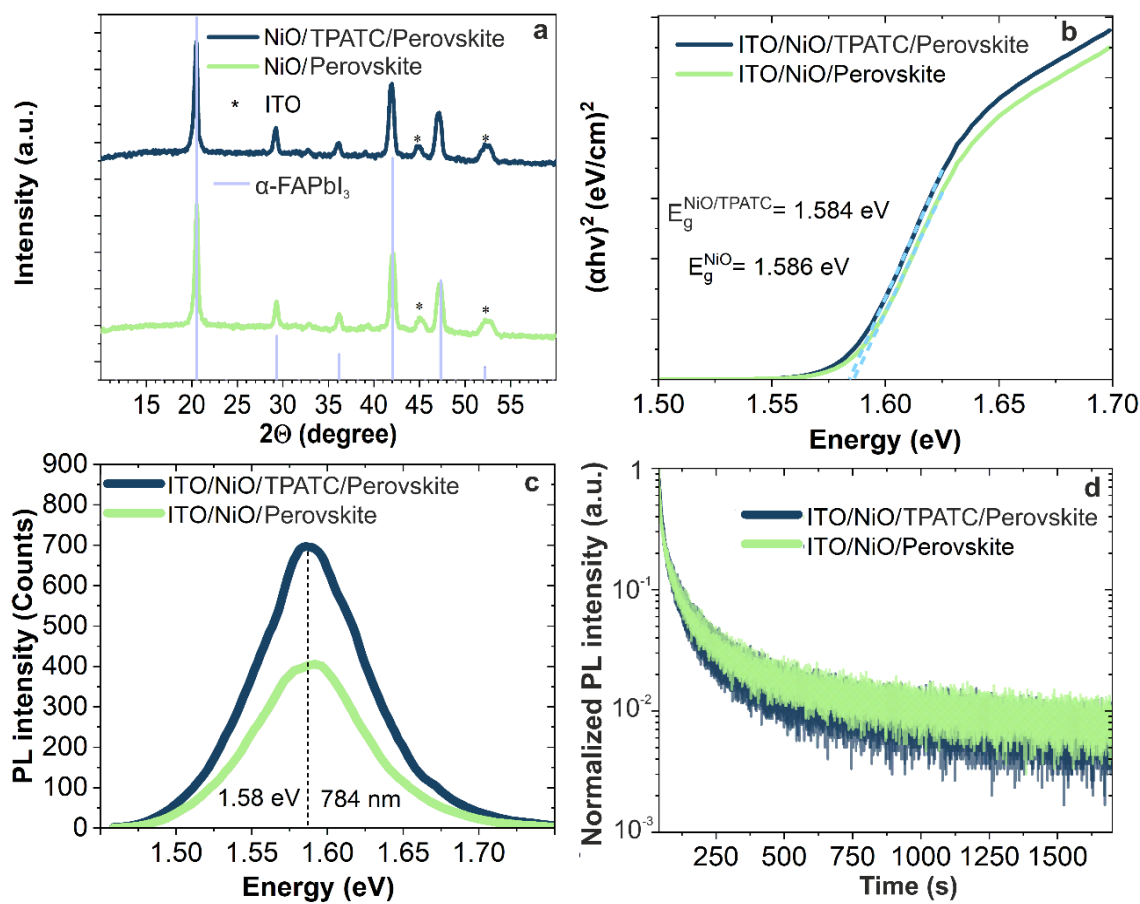

Figure 3 - Analysis of structural and optical properties of CsFAPbI$_{3-x}$Cl$_x$ perovskite deposited on ITO/NiO and ITO/NiO/TPATC HTL's; a) XRD; b) Tauc plot; c) PL; d) TRPL

PSCs with p-i-n architecture have been fabricated with the following architecture: ITO (300 nm)/NiO(hole transporting layer, 20 nm)/TPATC/perovskite absorber (450 nm)/C$_{60}$ (electron transporting layer, 40 nm)/bathocuproine (hole blocking layer, 8 nm)/Copper (100 nm) (See details in the Experimental Section). To investigate the output performance of PSCs fabricated with NiO (reference) and NiO/TPATC HTLs, we measured IV characteristics under standard illumination conditions of AAA class solar simulator (Xe lamp light source with AM 1.5G spectra filter, 1000 W/m$^2$). To simplify the naming of the PSCs in the manuscript in the following the device configuration will be attributed to the HTL (NiO and NiO/TPATC respectively). The statistical distribution for the calculated IV parameters, including open circuit voltage ($V_{OC}$), short circuit current density ($J_{SC}$), filling factor (FF), and power conversion efficiency (PCE), is presented in the box-charts on **Fig. S17** (ESI). The average values of $V_{OC}$ for PSCs increased from 1.041 V for NiO devices to 1.092 V for NiO/TPATC. Reference devices showed average values of $J_{SC}$ = 23.2 mA/cm$^2$, while a slight increase to 23.4 mA/cm$^2$ was observed for NiO/TPATC PSCs. Notably, the average fill factor for NiO/TPATC devices (75.5%) was smaller than in reference one (77.4%). Calculation of series resistance from IV curve showed an increase from 2.3 to 3.5 Ohm×cm$^2$ for PSCs with SAM. As a result, the PCE of the best (average) device was improved from 19.43% (18.5%) for the NiO PSCs to 20.58% (19.3%) for NiO/PSCs. The J-V characteristics for the champion devices are presented in **Fig.4(a)**. Best performing PSCs reached PCE=20.58% for the NiO/TPATC and 19.41% for the reference. The external quantum efficiency (EQE) spectra for the fabricated PSCs (**fig.4(b)**) reveals a slight improvement for the NiO/TPATC device, with the peak value of 91% at 600 nm. The integration of EQE spectra gave the $J_{SC}$ values of 22.9 mA/cm$^2$ for NiO TPATC and 22.1 mA/cm$^2$ for NiO devices, which are in agreement with data from IV curves. The stabilization of the output performance was estimated via maximum power point tracking measurements under standard illumination conditions (**fig.4(c)**). For reference, the power output ($P_{max}$) of perovskite solar cells (PSCs) showed minimal changes over time. The amplitude of these changes was no more than 2.5%, and after 600 seconds of measurements, the $P_{max}$ reached a gradual saturation. NiO/TPATC devices exhibited an initial rapid drop in $P_{max}$ tracking, followed by a slight increase in values during the measurements. The charge carrier extraction ability was analyzed via transient photocurrent measurements (TPC), as presented on the **fig. 4(d)**. When a solar cell is excited with a short pulse of light (see experimental for the details), the photogenerated charges are extracted on the electrodes, resulting in a current, which is detected during the measurements. Since the excitation pulse is square, there are two ways to measure TPC: in a "light on" and a "light off". TPC measurements yield information about dynamics of charge extraction and recombination processes. The obtained data represents the rising and falling profiles of the photocurrent under short circuit conditions. The rising and falling time values (**t$_r$** and **t$_f$**, respectively) were calculated as the period required to reach from 10% to 90% of the saturated photocurrent. The rise time ($t_r$) for the NiO PSC was 33 us, while the NiO/TPATC device showed a more rapid charge extraction dynamics with a $t_r$ of 18 us (45% smaller compared to the reference). The $t_f$ values show that the decay period of NiO devices (30 us) is 33% longer than that of NiO/TPATC samples ($t_f$ =20 us). It can be explained that TPATC SAM positively affects the HTL/absorber interface for the charge extraction efficiency. The hole collection junction modified with TPATC fills the traps under photo-injection conditions faster, indicating a reduced concentration of nonradiative recombination sites.

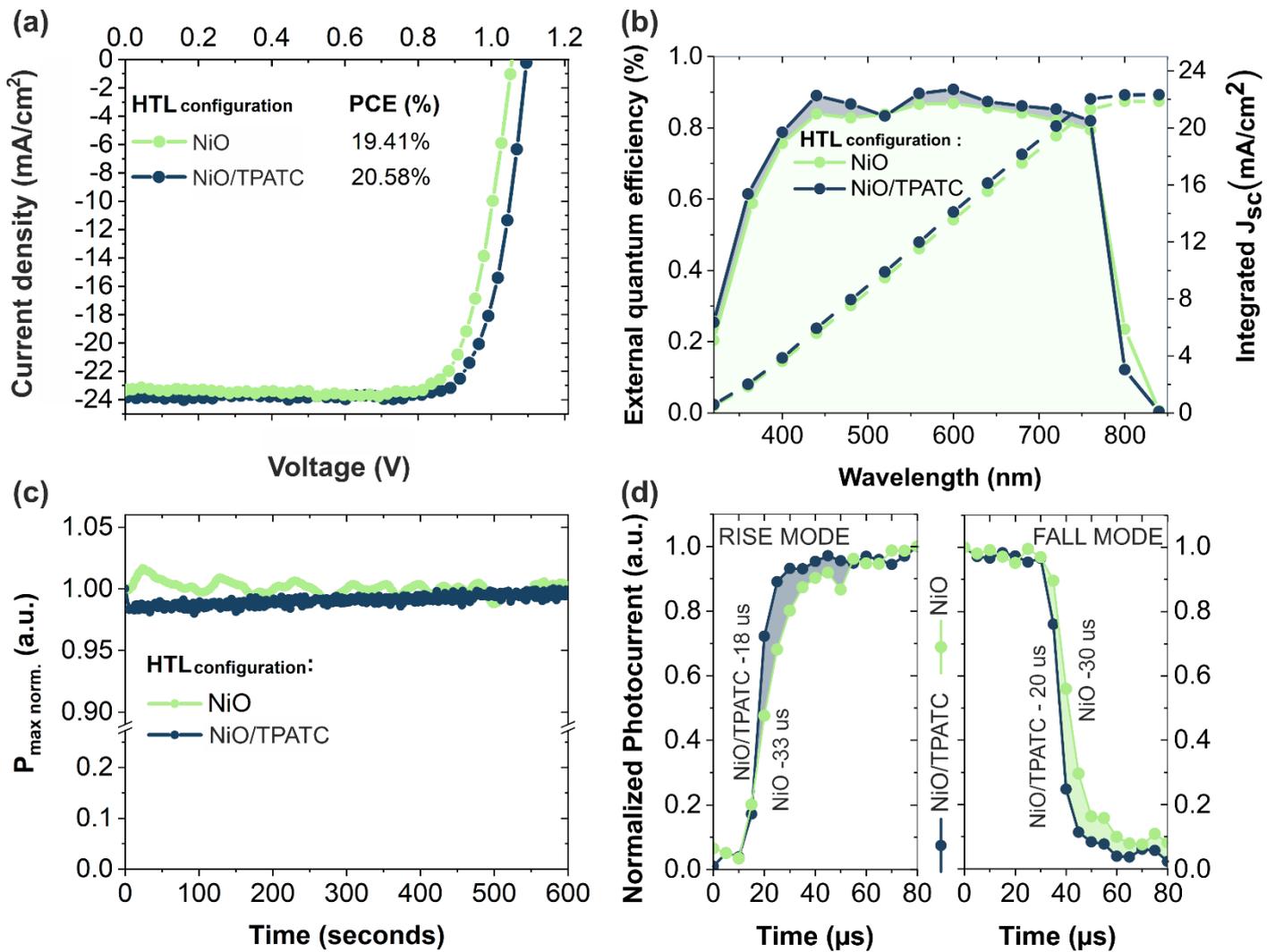

Figure 4 - The champion IV curves for PSCs with NiO and NiO/TPATC HTLs (a); External quantum efficiency spectra for PSCs PSCs with NiO and NiO/TPATC HTLs (b); Maximum power point tracking for PSCs under light soaking (c); Transient photo-current measurements for both device configurations

The increase in $V_{OC}$ and $J_{SC}$ were the main contributors to the increase in efficiency of NiO/TPATC perovskite solar cells (PSCs). The application of SAM at the hole-transport interface resulted in a complex outcome. The value of $V_{OC}$ in a solar cell is theoretically determined by the position of Fermi quasi-levels under photoinjection conditions, which also depends on the contribution of non-radiative recombination. The use of TPATC effectively increased the work function at the NiO/perovskite interface, optimizing the energy level alignment for hole collection. The improved lifetimes of the charge carriers (according to TRPL data) and increased intensity of the PL point to the partial passivation of trap states in the perovskite absorber.

The recognition of different light-soaking behavior indicates changes in the properties of the hole-transport interface, specifically the migration of ionic defects. The chemical instability and capacitive effects of the surface of NiO hole transport layers is a well-known phenomenon in p-i-n perovskite solar cells (PSCs). Defects in the structure of the NiO molecule can lead to the formation of uncoordinated $Ni^{3+}$ and $Ni^{2+}$ states. Oxidation of iodide at the NiO/perovskite interface can result in an electrochemical reaction, causing deprotonation of the organic cation A of the perovskite molecule[73]–[75]. This can negatively impact the collection of photo carriers and promote the accumulation/depletion processes.

Transport layer engineering and the study of its contact properties with the PSC active region remain open questions, as the influence of defects on this aspect is not fully understood. Recent studies[76]–[79] have attempted to address this issue using capacitance techniques, which are commonly employed for such problems in semiconductor physics, achieving quantitative control over defect parameters is crucial for analyzing the performance of semiconductor devices. Deep-Level Transient Spectroscopy (DLTS) and Admittance Spectroscopy (AS) measurements is widely regarded method for identification of the defect concentration, activation energies, diffusion coefficients, etc. in optoelectronic devices and particularly in PSCs. In this DLTS experiment, a steady-state voltage of -0.1 V and a 50 ms long filling pulse of +0.5 V were utilized, while the opposite biasing conditions (+0.5 V as steady-state and -0.1 V pulse) were employed for the so-called RDLTS. AS data were collected at 0 V over a frequency range of 0.02 to 1000 kHz. Defect concentrations were estimated through capacitance screening at the Debye length, activation energies and diffusion coefficients were determined using the Arrhenius plot (see supplementary materials).

In order to estimate the impact of the interface engineering using TPATC we analyzed the changes in DLTS spectra (**fig.S18,** ESI) for the fabricated devices. In our work we used hybrid chemical composition of the perovskite absorber, thus, the analysis of the obtained data requires considering of $FAPbI_3$ and $CsPbI_3$ perovskites. The calculated defect parameters presented in the **table 3** for the results of admittance and DLTS measurements. Calculated defect concentrations ($N_t$) show a decrease in magnitude for NiO/TPATC samples. The data extracted from admittance spectroscopy was $1.0×10^{15}$ cm$^{-3}$ for the NiO sample, which was approximately ten times larger compared to NiO/TPATC device with $N_t=7.2×10^{14}$ cm$^{-3}$. We observed a similar relationship in the variation of concentrations from R-DLTS and DLTS measurement. However, the difference in $N_t$ between the DLTS data for NiO PSCs and that for NiO/TPATC devices was less than one order of magnitude.

Table 3 – Defects concentration, activation energies and diffusion coefficients for NiO and NiO/TPATC PSCs

| Sample | Method | Concentration (cm$^{-3}$) | Activation energy (eV) | Diffusion Coefficient at 300K (cm$^2$ s$^{-1}$) | Possible origin [Link] |
|---|---|---|---|---|---|
| NiO | Admittance | 1.0×10$^{15}$ | 0.509 | 1.7×10$^{-7}$ | $V_{FA}$ or $V_I$ [72][73] [74] |
|  | DLTS | 2.5×10$^{12}$ | 0.646 | 1.3×10$^{-8}$ | $V_I$ or $I_i$ [72][74] |
|  | R-DLTS | 1.8×10$^{14}$ | 0.451 | 7.7×10$^{-9}$ | $V_I$[74] |
| NiO/TPATC | Admittance | 7.2×10$^{14}$ | 0.457 | 3.9×10$^{-7}$ | $V_I$[74] |
|  | DLTS | 1.4×10$^{12}$ | 0.766 | 3.3×10$^{-8}$ | $I_{Pb}$ [75] |
|  | R-DLTS | 1.3×10$^{13}$ | 0.643 | 7.7×10$^{-9}$ | $V_I$ or $I_i$ [72][74] |

The admittance spectroscopy results showed that the defect activation energy for NiO devices was 0.509 eV, while for NiO/TPATC configuration, it was 0.457 eV. Accord to the literature[80], [81] , the energy of 0.509 eV may correspond to FA vacancy ($V_{FA}$), while the energy of 0.457 eV is close with formation of iodine vacancy ($V_I$)[82]. DLTS measurements showed defects of 0.646 and 0.766 eV for NiO and NiO/TPATC samples, respectively. The ~0.6 eV defects could be related also for the $V_I$[82] or for the iodine interstitials ($I_i$)[80]. The 0.766 eV defect can be identified as iodine-lead antisite ($I_{Pb}$)[83]. R-DLTS data demonstrated the presence of 0.451 eV ($V_I$) peak for NiO PSC and 0.643 eV peal for NiO/TPATC sample. Interestingly, that the signatures of ~0.6 eV defects were measured for NiO and NiO/TPATC samples under opposite bias conditions. The TPATC interlayer may play a role in neutralizing certain types of charged ionic defects, as suggested by observed evidence. Migration of such defects in PSC structures can be attributed to parasitic effects of photocarrier trapping, as well as the accumulation of ion-clusters that form potential barriers at the interfaces[84]. The application of various methods for electrical characterization of defects in PSCs provided comprehensive information about their properties under different conditions. We identified the matching in levels with activation energies of ~0.45 eV ($V_I$) and ~0.64 eV ($V_I$ or $I_i$) for both configurations of the investigated samples. We also observed clear differences in the presence of $V_{FA}$ defects for NiO devices and signals I-Pb antisites for NiO/TPATC. It is worth noting that comparing the results obtained by different methods is very complicated due to the sensitivity of device structures based on halide perovskites to the conditions and type of the measurements[85]. According to reports[86]–[88], A-site vacancy defects in perovskite molecule can be formed due to the formation of volatile products, such as formamidine. The migration of $V_{FA}$ defects initiates local band bending[89], which reduces the efficiency of charge carrier collection. At the same time, I-Pb antisites have the potential to function as deep non-radiative recombination centers[90]. For both device types, the levels associated with X-anion position defects in the perovskite molecule have also been determined. In p-i-n PSCs, iodine interstitials can participate in trapping of photocarriers[91], while positively charged $V_I$ migrates to the cathode, inducing unfavorable accumulation effects[92]. Thus, it can be concluded that the application of interface engineering with TPATC has a complex effect on the nature of defects in p-i-n PSCs. Modification of the NiO surface allows the reconfiguration of defect levels in $CsFAPbI_{3-x}Cl_x$ perovskite and reduces the concentration of traps. For NiO/TPATC, imperfections associated with the A-site organic cation are suppressed, but all types of defects were related to the iodine anion (vacancies, antisites, interstitials).

Besides improving the power conversion efficiency of the devices with TPATC, the interface engineering provided the stabilization of the output performance under photo-stress conditions (LED illumination, refer to the Experimental section for details). The NiO and NiO/TPATC PSCs were encapsulated and tested in ambient air conditions. We estimated stability of the device operation at open circuit conditions under continuous light soaking (LS) following the ISOS-L-2 protocol[93]. The temperature of the samples in the testing setup was 63±1.5ºC. The stability of PSCs was tested during 1000 h (**fig.5 (a)**). Analysis of the maximum power stability ($P_{max}$) measurement for champion devices shows that for both device configurations, the values increase by 3-5% during the initial period of light soaking. For NiO PSCs, a smooth increase in $P_{max}$ was observed for 200 hours, while for NiO/TPATC devices, the same period was ~50 hours. The subsequent change in $P_{max}$ values for NiO samples showed a slight downward trend to 92% after 1000 hours of testing. NiO/TPATC devices exhibited slight fluctuations near a plateau of values at 98% of the initial $P_{max}$. Device stabilization statistics are an important indicator of the relevance of the tests performed. In our work, we made stability analysis on a batch of 10 devices for each of their configurations. The data presented in Figure 5 clearly shows that NiO PSCs have a strong variation in values and lose more than 20-30% of the initial $P_{max}$ in the first 400 hours of testing. In contrast, NiO/TPATC devices show high reproducibility in stabilizing the maximum power over 1150 hours.

In parallel to the analysis of the photoelectric parameters, we have investigated the phase composition changes in multilayer device structures under light soaking conditions (high intensity LED illumination and heating to 63.5 °C). For the measurements we used the XRD method, the samples were fabricated in simplified multilayer stacks on ITO glass substrates with HTL/perovskite absorber/ETL structure without a metallic electrode and encapsulation. The samples were exposed to light and heating for 1000 hours in a glove box with an inert nitrogen environment. **Fig.5 (b-c)** illustrate the evolution of the phase composition of the NiO and NiO/TPATC configurations for "as fabricated" conditions and "after 1000 hours" of light soaking. As fabricated samples of both types contained α-$FAPbI_3$ (the main phase of the photoactive layer) and $In_{1.88}Sn_{0.12}O_3$ (the ITO substrate signal) phases without significant differences. After prolonged exposure to light soaking stress, the XRD spectra for NiO samples showed a peak at 18.5°, as well as a large number of weakly distinguishable low-intensity peaks: 14.7°, 31.5°, 34°, 37-42°, 50°, and 55°. The appearance of new peaks indicates that the perovskite film has undergone phase decomposition, resulting in the separation of the crystalline phase of lead iodide oriented along the 001 direction and the inorganic perovskite $CsPbI_3$. For the NiO/TPATC sample, we determined the appearance of $CsPbI_3$ phase peaks of low intensity. In contrast to NiO sample, we observed a distinct difference in the phase decomposition of perovskite on the NiO/TPATC interlayer: the absence of the lead iodide peak at 18.5° after 1000h of light soaking. Therefore, we can conclude that the use of TPATC interlayer for NiO HTL can improve the resistance of perovskite absorbers to the formation of 001 $PbI_2$ phase under prolonged exposure to environmental stressors. Changes in the phase composition of the perovskite absorber under external degradation factors are related to the nature of defects formed during the process of device fabrication. According to many reports in the literature[94]–[96], perovskite molecule decomposition and phase segregation can be triggered through chemical interaction with ionic defects[97]. For NiO devices, defects related to the A-cation position were inhibited during the device fabrication stage and were not the result of external influences. Initially, NiO PSCs didn't have a complete stoichiometric balance between A (FA) and B (Pb) position cations due to the presence of point defects in the perovskite absorber. Therefore, the multilayer structures on NiO HTL prone to form the $PbI_2$ phase caused by FA deficiency. Long-term exposure to heating and light-soaking accelerated the formation and accumulation of $V_{FA}$ defect clusters, which eventually led to the appearance of $PbI_2$, indicative of the decomposition process. Interface engineering with TPATC suppressed formation imperfections associated with the organic cation, which provided stabilization of the phase composition and more stable operation of the solar cells.

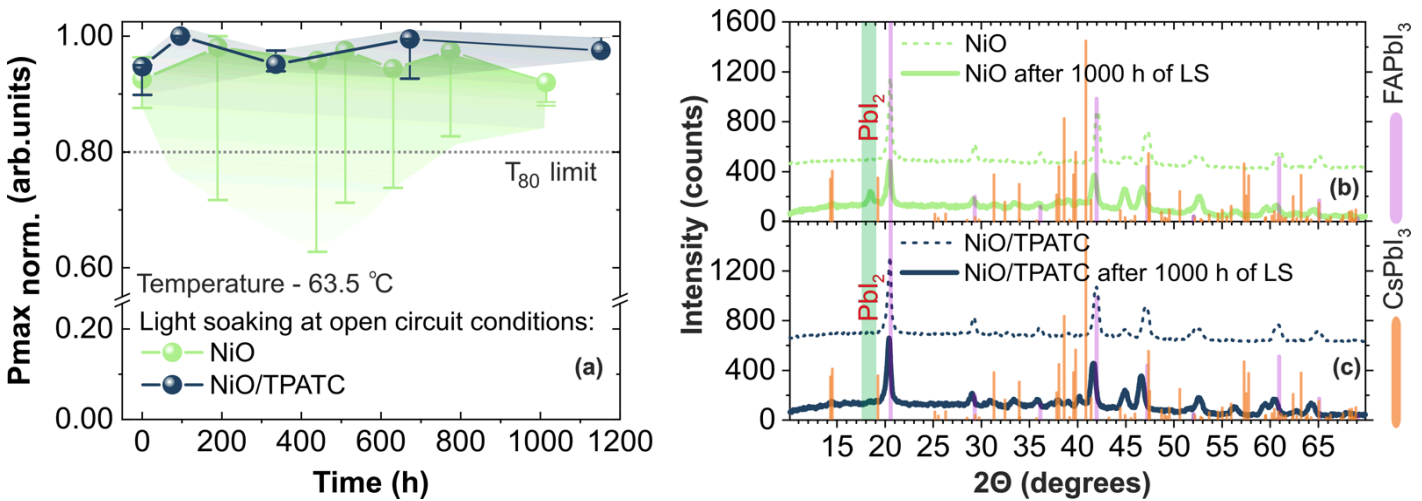

Figure 5 – The stability of $P_{max}$ values for the PSCs in NiO and NiO/TPATC configuration (a). Points referred to the best performing samples and colored background referred to the standard deviation of the values calculated for the batch of the devices. Evolution of the XRD spectra for NiO PSCs under LS stress (b); Evolution of the XRD spectra for NiO/TPATC PSCs under LS stress (c)

To estimate the potential of interface engineering with TPATC SAM for the up-scaling, we fabricated perovskite solar modules (PSMs, photo-image presented on the **fig6(a)**) with an active area of 64.8 cm$^2$ (12 sub-cells) on glass substrates (100x100 mm$^2$). The PSMs were fabricated with in-series connection of the sub-cells using laser-scribing patterning accord to the work of Castriotta[98]. The width of each sub-cell's active area was 6.503 mm, and the width of "dead" interconnection area was 480 microns. TPATC interlayer was slot-die coated on the pre-patterned ITO/NiO substrates in ambient conditions (see Experimental section for the details of PSM's fabrication). Champion IV performance measured at standard 1.5 AM G illumination for NiO and NiO/TPATC PSMs are shown on the **fig.6(b)** and **tab.4**. Statistical distribution of the output IV parameters for the set of PSMs presented in box-charts of the **fig. S19** (ESI). The use of TPATC interlayer for perovskite solar modules showed a comprehensive enhancement of all IV characteristics.[98] For the best modules of NiO and NiO/TPATC, PCE increased from 13.33% to 15.64%. The main impact for the improvement was given by gained $V_{oc}$ and FF values. The short circuit current values in NiO/TPATC PSMs slightly increased from 109 to 114 mA.

Table 4 – Output IV performance of the fabricated PSMs

| PSM configuration | $V_{oc}$ (V) | $I_{sc}$ (mA) | FF (%) | $P_{max}$ (mW) | PCE (%) |
|---|---|---|---|---|---|
| NiO | 11.938 | 109.36 | 66 | 861.07 | 13.33 |
| NiO/TPATC | 12.473 | 114.15 | 71 | 1012.23 | 15.64 |

The impact of shunt effects[99] and non-radiative losses[100] strongly manifests[100] in the operation of solar cells at low-light conditions when concentration of the photo-injected charge carriers is significantly reduced in comparison to standard 1 sun illumination. Potentially, photocarrier trapping at low-light intensities[101]–[104] in large-scale perovskite solar modules can critically affect the power output during morning, evening, or under shading conditions. To analyze the impact of HTL/perovskite interface properties on device performance under low-light conditions, we measured the IV characteristics of PSMs at outdoor ambient illumination with light intensity ($I_L$) ranging from ~30 to ~18500 Lx (see photo-images on **fig.6 (c),(d)**). NiO/TPATC PSM exhibited $V_{oc}$= 11.67 V at 13600 Lx, while reference PSM showed only 9.87 V even at higher intensity of 15000 Lx (**fig.6(e)**). At ultralow-light conditions with $I_L$~30 Lx, the $V_{oc}$ for NiO/TPATC PSM was 7.48 V and for the NiO PSM the value dramatically decreased to 3.94 V. Relative comparison of the $V_{oc}$ values at low-light ambient conditions shows that: at $I_L$ ~10$^4$ Lx the use of TPATC gives an increase of 15.4%; at $I_L$~10$^2$ Lx, the increase reaches 17.7%; at $I_L$~10$^1$ the increase is 47.3%. The analysis of $I_{sc}$ versus $I_L$ demonstrated that both module configuration have almost similar linear dependence with slight improvement for NiO/TPATC device (**fig.S20** in ESI). Comparison of the FF for PSMs under low-light conditions revealed improved behavior for NiO/TPATC. For both configurations of PSMs, a nonlinear trend of fill factor was observed with changes in light intensity. Notably, we observed a peak FF value at $I_L$~ 5000 Lx. At an $I_L$ of 5000 lux, the maximum fill factor (FF) value for NiO PSMs was 71%, while for NiO/TPATC PSMs, it was 64%. Reducing the light intensity to 30 lux resulted in a decrease in the FF to 58% for NiO/TPATC PSMs and to 50% for NiO PSMs. Total contribution of IV parameters was estimated through the calculation of maximum power ($P_{max}$) values (see **fig. 6(f)**). At an illumination of $10^4$ lux, the NiO/TPATC device exhibited a remarkable 47% increase in maximum $P_{max}$ compared to the NiO PSM device. This increase further improved to 90% at an ultra-low illumination level of $10^1$ lux.

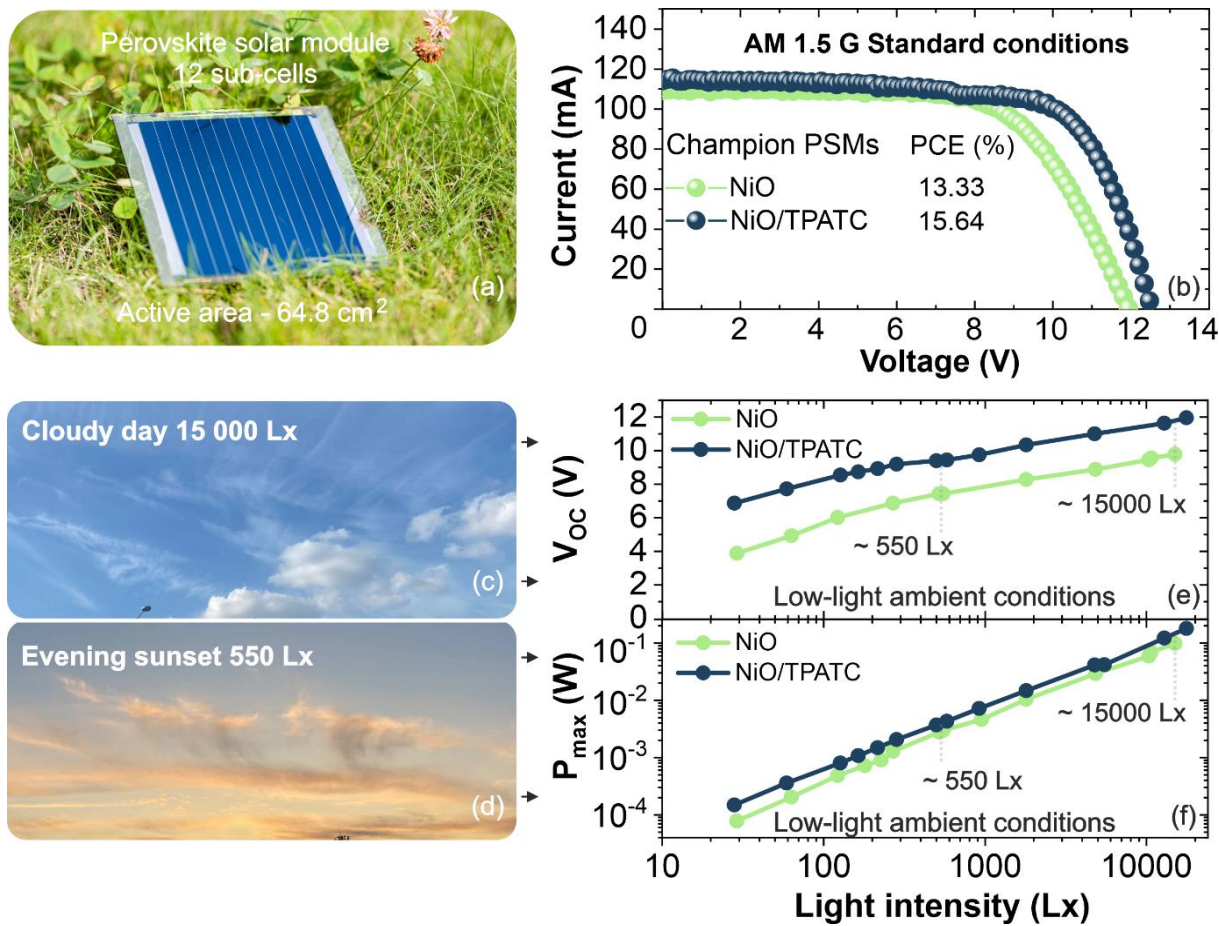

Figure 6 – Perovskite solar module fabricated with TPATC interlayer (a); champion IV curves for the PSMs fabricated in NiO and NiO/TPATC configurations respectively (b); photo images of the ambient illumination conditions with corresponding light intensities 15 000 Lx (c) and 550 Lx (d); The dependencies of $V_{oc}$ and $P_{max}$ values measured for PSMs at different conditions of low-light ambient illumination

**Conclusions**

We synthesized a new self-assembled monolayer HTM 5-[4-(diphenylamino)phenyl]thiophene-2-carboxylic acid (TPATC) in two steps from commercially available materials. In addition to the simple synthesis, the undoubted advantages of TPATC such as high thermal stability, suitable HOMO level for the hole transport from perovskite and absence of parasitic absorption relative to the active layer were revealed, which make it very promising for tuning of hole-collection interface in p-i-n PSCs. Surface modification of a non-stoichiometric NiO thin-film with an interlayer TPATC (4 nm) reduced the contribution of non-radiative recombination processes at the heterojunction with a multicationic perovskite absorber ($CsFAPbI_{3-x}Cl_x$). The application of TPATC rearranged the energy level alignment between NiO and CsFAPbI$_{3-x}$Cl$_x$, resulting in an increase in the HTL's work function value by ~0.2 eV. IV performance measurements under standard AM 1.5 G conditions for small area devices (0.15 cm$^2$) showed a confirmed improvement in charge collection efficiency for samples with modified interfaces. The V$_{oc}$ values for PSCs increased from 1.041 V for NiO devices to 1.092 V for NiO/TPATC. Meanwhile, the PCE of champion samples improved from 19.43% for references to 20.58% for devices with TPATC. Measurements by admittance and deep-level transient spectroscopies allowed to gather numerical data for ionic defects. Calculations showed that modifying the HTL interface with TPATC reduces the trap concentration by about one order of magnitude relative to the NiO reference configuration. For NiO PSCs, we observed the energy levels of 0.509 eV, 0.646 eV, and 0.451 eV corresponding to V$_{FA}$, I$_i$, and V$_I$, respectively. NiO/TPATC devices showed close values of V$_I$ (0.457 eV) and I$_i$ (0.606 eV), as well as deeper antisite I$_{Pb}$ (0.766 eV) spectral features. We estimated the stability of the device operation at open circuit conditions under continuous light soaking (LS) following the ISOS-L-2 protocol. NiO/TPATC devices showed slight fluctuations near a plateau at 98% with respect to the initial value. Analysis of the statistical distribution of power loss data for a group of devices clearly determined that the reference NiO PSCs exhibit losses of more than 20% after 400 hours of light-soaking. The phase composition of multilayer device stacks under external stress evolved differently depending on the interface modification. Samples with TPATC were found to be more resistant to the formation of the PbI$_2$ phase after 1000 h of light-soaking compared to those with NiO. The interface modification with TPATC showed the beneficial effect for up-scaled perovskite solar modules (12 sub-cells) with an active area of 64.8 cm$^2$. Application of slot-die-coated TPATC increased the PCE at AM 1.5 G conditions from 13.22% for NiO PSM to 15.64% and boosted the performance under ambient low-light conditions with intensity from 10$^1$ to 10$^4$ Lx. Maximum power for NiO/TPATC modules was 47-90% higher with respect to NiO samples across the entire range of low-light illumination intensities.

The modification of HTL/perovskite interface with TPATC has shown a complex effect on the transport of charge carriers in perovskite solar cells and the degradation dynamics under the influence of external factors. Such features as good energy level alignment and strong interaction of TPATC with NiO coupled with the hydrophobic nature of NiO/TPATC surface allowed to obtain less defective film of CsFAPbI$_{3-x}$Cl$_x$ with decreased defect density. This contributed to their increased phase stability and resistance for defect-induced decomposition. The development of this approach for the ETL/ perovskite interface may be further expanded for layers of stable n-type oxide semiconductors such as SnO$_2$ and ZnO nanoparticles.

**Acknowledgments**

The authors gratefully acknowledge the financial support from Russian Science Foundation with project № 22-19-00812.

# References

[1] Z.-K. Tang, Z.-F. Xu, D.-Y. Zhang, S.-X. Hu, W.-M. Lau, and L.-M. Liu, "Enhanced optical absorption via cation doping hybrid lead iodine perovskites," *Sci Rep*, vol. 7, no. 1, p. 7843, 2017, doi: 10.1038/s41598-017-08215-3.

[2] E. Alarousu *et al.*, "Ultralong Radiative States in Hybrid Perovskite Crystals: Compositions for Submillimeter Diffusion Lengths," *Journal of Physical Chemistry Letters*, vol. 8, no. 18, pp. 4386–4390, Sep. 2017, doi: 10.1021/ACS.JPCLETT.7B01922.

[3] H. Oga, A. Saeki, Y. Ogomi, S. Hayase, and S. Seki, "Improved understanding of the electronic and energetic landscapes of perovskite solar cells: High local charge carrier mobility, reduced recombination, and extremely shallow traps," *J Am Chem Soc*, vol. 136, no. 39, pp. 13818–13825, 2014, doi: 10.1021/ja506936f.

[4] T. Leijtens *et al.*, "Electronic properties of meso-superstructured and planar organometal halide perovskite films: Charge trapping, photodoping, and carrier mobility," *ACS Nano*, vol. 8, no. 7, pp. 7147–7155, 2014, doi: 10.1021/nn502115k.

[5] M. Grätzel, "The light and shade of perovskite solar cells," *Nature Publishing Group*, vol. 13, no. 9, pp. 838–842, 2014, doi: 10.1038/nmat4065.

[6] "National Renewable Energy Laboratory (NREL) Home Page | NREL."

[7] T. S. Le *et al.*, "All-Slot-Die-Coated Inverted Perovskite Solar Cells in Ambient Conditions with Chlorine Additives," *Solar RRL*, vol. 6, no. 2, p. 2100807, Feb. 2022, doi: 10.1002/SOLR.202100807.

[8] F. Matteocci *et al.*, "Fabrication and Morphological Characterization of High-Efficiency Blade-Coated Perovskite Solar Modules," *ACS Appl Mater Interfaces*, 2019, doi: 10.1021/acsami.9b05730.

[9] M. Cai, Y. Wu, H. Chen, X. Yang, Y. Qiang, and L. Han, "Cost-Performance Analysis of Perovskite Solar Modules," *Advanced Science*, 2017, doi: 10.1002/advs.201600269.

[10] S. P. Dunfield *et al.*, "From Defects to Degradation: A Mechanistic Understanding of Degradation in Perovskite Solar Cell Devices and Modules," *Adv Energy Mater*, vol. 10, no. 26, p. 1904054, Jul. 2020, doi: 10.1002/aenm.201904054.

[11] M. Alsari *et al.*, "Degradation Kinetics of Inverted Perovskite Solar Cells," *Sci Rep*, vol. 8, no. 1, p. 5977, Dec. 2018, doi: 10.1038/s41598-018-24436-6.

[12] J. Bisquert, E. J. J. Orcid, and E. J. J. Orcid, "The Causes of Degradation of Perovskite Solar Cells," *J Phys Chem Lett*, vol. 10, pp. 5889–5891, 2019, doi: 10.1021/acs.jpclett.9b00613.

[13] N. Yaghoobi Nia, D. Saranin, A. L. Palma, and A. Di Carlo, "Perovskite solar cells," in *Solar Cells and Light Management*, Elsevier, 2020, pp. 163–228. doi: 10.1016/B978-0-08-102762-2.00005-7.

[14] C. Manspeaker, S. Venkatesan, A. Zakhidov, and K. S. Martirosyan, "Role of interface in stability of perovskite solar cells," *Curr Opin Chem Eng*, vol. 15, pp. 1–7, Feb. 2017, doi: 10.1016/J.COCHE.2016.08.013.

[15] T. H. Han, S. Tan, J. Xue, L. Meng, J. W. Lee, and Y. Yang, "Interface and Defect Engineering for Metal Halide Perovskite Optoelectronic Devices," *Advanced Materials*, vol. 31, no. 47, pp. 1–35, 2019, doi: 10.1002/adma.201803515.

[16] D. Di Girolamo *et al.*, "From Bulk to Surface: Sodium Treatment Reduces Recombination at the Nickel Oxide/Perovskite Interface," *Adv Mater Interfaces*, vol. 6, no. 17, p. 1900789, Sep. 2019, doi: 10.1002/admi.201900789.

[17] G. Grancini *et al.*, "One-Year stable perovskite solar cells by 2D/3D interface engineering," *Nat Commun*, vol. 8, p. 15684, Jun. 2017, doi: 10.1038/ncomms15684.

[18]  A. F. Akbulatov *et al.*, "When iodide meets bromide: Halide mixing facilitates the light-induced decomposition of perovskite absorber films," *Nano Energy*, vol. 86, p. 106082, Aug. 2021, doi: 10.1016/j.nanoen.2021.106082.

[19]  W. C. Lin, W. C. Lo, J. X. Li, Y. K. Wang, J. F. Tang, and Z. Y. Fong, "In situ XPS investigation of the X-ray-triggered decomposition of perovskites in ultrahigh vacuum condition," *Npj Mater Degrad*, vol. 5, no. 1, Dec. 2021, doi: 10.1038/S41529-021-00162-9.

[20]  E. J. Juarez-Perez, L. K. Ono, M. Maeda, Y. Jiang, Z. Hawash, and Y. Qi, "Photodecomposition and thermal decomposition in methylammonium halide lead perovskites and inferred design principles to increase photovoltaic device stability," *J Mater Chem A Mater*, 2018, doi: 10.1039/c8ta03501f.

[21]  M. F. N. Taufique, R. Khanal, S. Choudhury, and S. Banerjee, "Impact of iodine antisite (IPb) defects on the electronic properties of the (110) CH3NH3PbI3 surface," *J Chem Phys*, vol. 149, no. 16, p. 164704, Oct. 2018, doi: 10.1063/1.5044667.

[22]  W. Li, J. Liu, F. Q. Bai, H. X. Zhang, and O. V. Prezhdo, "Hole Trapping by Iodine Interstitial Defects Decreases Free Carrier Losses in Perovskite Solar Cells: A Time-Domain Ab Initio Study," *ACS Energy Lett*, 2017, doi: 10.1021/acsenergylett.7b00183.

[23]  G. Sadoughi *et al.*, "Observation and Mediation of the Presence of Metallic Lead in Organic-Inorganic Perovskite Films," *ACS Appl Mater Interfaces*, vol. 7, no. 24, pp. 13440–13444, Jun. 2015, doi: 10.1021/ACSAMI.5B02237/SUPPL_FILE/AM5B02237_SI_001.PDF.

[24]  J. S. Park, J. Calbo, Y. K. Jung, L. D. Whalley, and A. Walsh, "Accumulation of Deep Traps at Grain Boundaries in Halide Perovskites," *ACS Energy Lett*, 2019, doi: 10.1021/acsenergylett.9b00840.

[25]  B. Chen, P. N. Rudd, S. Yang, Y. Yuan, and J. Huang, "Imperfections and their passivation in halide perovskite solar cells," *Chemical Society Reviews*. 2019. doi: 10.1039/c8cs00853a.

[26]  M. Can *et al.*, "Electrical properties of SAM-modified ITO surface using aromatic small molecules with double bond carboxylic acid groups for OLED applications," *Appl Surf Sci*, vol. 314, pp. 1082–1086, Sep. 2014, doi: 10.1016/j.apsusc.2014.05.181.

[27]  F. Ali, C. Roldán-Carmona, M. Sohail, and M. K. Nazeeruddin, "Applications of Self-Assembled Monolayers for Perovskite Solar Cells Interface Engineering to Address Efficiency and Stability," *Adv Energy Mater*, vol. 10, no. 48, Dec. 2020, doi: 10.1002/aenm.202002989.

[28]  A. Ulman, "Formation and Structure of Self-Assembled Monolayers," *Chem Rev*, vol. 96, no. 4, pp. 1533–1554, Jan. 1996, doi: 10.1021/cr9502357.

[29]  Y. Lin *et al.*, "Self-Assembled Monolayer Enables Hole Transport Layer-Free Organic Solar Cells with 18% Efficiency and Improved Operational Stability," *ACS Energy Lett*, vol. 5, no. 9, pp. 2935–2944, Sep. 2020, doi: 10.1021/acsenergylett.0c01421.

[30]  W. Cai, Z. Zang, and L. Ding, "Self-assembled monolayers enhance the performance of oxide thin-film transistors," *Journal of Semiconductors*, vol. 42, no. 3, p. 030203, Mar. 2021, doi: 10.1088/1674-4926/42/3/030203.

[31]  L. A. Frolova *et al.*, "Advanced Nonvolatile Organic Optical Memory Using Self-Assembled Monolayers of Porphyrin–Fullerene Dyads," *ACS Appl Mater Interfaces*, vol. 14, no. 13, pp. 15461–15467, Apr. 2022, doi: 10.1021/acsami.1c24979.

[32]  Q. Jiang *et al.*, "Towards linking lab and field lifetimes of perovskite solar cells," *Nature*, Sep. 2023, doi: 10.1038/s41586-023-06610-7.

[33]  A. Farag *et al.*, "Evaporated Self-Assembled Monolayer Hole Transport Layers: Lossless Interfaces in *p-i-n* Perovskite Solar Cells," *Adv Energy Mater*, vol. 13, no. 8, Feb. 2023, doi: 10.1002/aenm.202203982.

[34]  Q. Liao *et al.*, "Self-assembled donor-acceptor hole contacts for inverted perovskite solar cells with an efficiency approaching 22%: The impact of anchoring groups," *Journal of Energy Chemistry*, vol. 68, pp. 87–95, May 2022, doi: 10.1016/j.jechem.2021.11.001.

[35] L.-L. Tan *et al.*, "Novel organic dyes incorporating a carbazole or dendritic 3,6-diiodocarbazole unit for efficient dye-sensitized solar cells," *Dyes and Pigments*, vol. 100, pp. 269–277, Jan. 2014, doi: 10.1016/j.dyepig.2013.09.025.

[36] E. Yalcin *et al.*, "Semiconductor self-assembled monolayers as selective contacts for efficient PiN perovskite solar cells," *Energy Environ Sci*, vol. 12, no. 1, pp. 230–237, 2019, doi: 10.1039/C8EE01831F.

[37] S. Liu, V. P. Biju, Y. Qi, W. Chen, and Z. Liu, "Recent progress in the development of high-efficiency inverted perovskite solar cells," *NPG Asia Mater*, vol. 15, no. 1, p. 27, May 2023, doi: 10.1038/s41427-023-00474-z.

[38] S. Maenosono, Y. Yamaguchi, and K. Yoshi, "Thin films of semiconductor nanocrystals self-assembled by wet coating," in *Studies in Surface Science and Catalysis*, Elsevier Inc., Jan. 2001, pp. 533–536. doi: 10.1016/s0167-2991(01)82146-7.

[39] W. Chen *et al.*, "Molecule-Doped Nickel Oxide: Verified Charge Transfer and Planar Inverted Mixed Cation Perovskite Solar Cell," *Advanced Materials*, 2018, doi: 10.1002/adma.201800515.

[40] D. Soo Kim and H. Chul Lee, "Nickel vacancy behavior in the electrical conductance of nonstoichiometric nickel oxide film," *J Appl Phys*, vol. 112, no. 3, p. 034504, Aug. 2012, doi: 10.1063/1.4742993.

[41] D. Soo Kim and H. Chul Lee, "Nickel vacancy behavior in the electrical conductance of nonstoichiometric nickel oxide film," *J Appl Phys*, vol. 112, no. 3, p. 034504, Aug. 2012, doi: 10.1063/1.4742993.

[42] X. Yin, Y. Guo, H. Xie, W. Que, and L. B. Kong, "Nickel Oxide as Efficient Hole Transport Materials for Perovskite Solar Cells," *Solar RRL*, vol. 3, no. 5, p. 1900001, May 2019, doi: 10.1002/solr.201900001.

[43] A. B. Huang *et al.*, "Achieving high-performance planar perovskite solar cells with co-sputtered Co-doping NiO $_x$ hole transport layers by efficient extraction and enhanced mobility," *J Mater Chem C Mater*, vol. 4, no. 46, pp. 10839–10846, Nov. 2016, doi: 10.1039/C6TC03624D.

[44] Y. Du *et al.*, "Polymeric Surface Modification of NiOx-Based Inverted Planar Perovskite Solar Cells with Enhanced Performance," *ACS Sustain Chem Eng*, 2018, doi: 10.1021/acssuschemeng.8b04078.

[45] D. Saranin *et al.*, "Transition metal carbides (MXenes) for efficient NiO-based inverted perovskite solar cells," *Nano Energy*, 2021, doi: 10.1016/j.nanoen.2021.105771.

[46] W. Chen, F. Z. Liu, X. Y. Feng, A. B. Djurišić, W. K. Chan, and Z. B. He, "Cesium Doped NiOxas an Efficient Hole Extraction Layer for Inverted Planar Perovskite Solar Cells," *Adv Energy Mater*, vol. 7, no. 19, 2017, doi: 10.1002/aenm.201700722.

[47] H. S. Kim *et al.*, "Effect of cs-incorporated NiOx on the performance of perovskite solar cells," *ACS Omega*, 2017, doi: 10.1021/acsomega.7b01179.

[48] D. Di Girolamo *et al.*, "Progress, highlights and perspectives on NiO in perovskite photovoltaics," *Chemical Science*. 2020. doi: 10.1039/d0sc02859b.

[49] L.-L. Tan *et al.*, "Novel organic dyes incorporating a carbazole or dendritic 3,6-diiodocarbazole unit for efficient dye-sensitized solar cells," *Dyes and Pigments*, vol. 100, pp. 269–277, Jan. 2014, doi: 10.1016/j.dyepig.2013.09.025.

[50] E. Yalcin *et al.*, "Semiconductor self-assembled monolayers as selective contacts for efficient PiN perovskite solar cells," *Energy Environ Sci*, vol. 12, no. 1, pp. 230–237, 2019, doi: 10.1039/C8EE01831F.

[51] H. D. Pham, T. C. Yang, S. M. Jain, G. J. Wilson, and P. Sonar, "Development of Dopant-Free Organic Hole Transporting Materials for Perovskite Solar Cells," *Adv Energy Mater*, vol. 10, no. 13, Apr. 2020, doi: 10.1002/aenm.201903326.

[52] W. Li, M. Huang, Y. Li, Z. Yang, and J. Qu, "Novel D-π-A-D Type aggregation induced emission luminogens based nanoparticles as efficient 1O2 photosensitizer and bright NIR imaging agent," *Dyes and Pigments*, vol. 186, p. 109041, Feb. 2021, doi: 10.1016/j.dyepig.2020.109041.

[53] M. D. Irwin, D. B. Buchholz, A. W. Hains, R. P. H. Chang, and T. J. Marks, "*p*-Type semiconducting nickel oxide as an efficiency-enhancing anode interfacial layer in polymer bulk-heterojunction solar cells," *Proceedings of the National Academy of Sciences*, vol. 105, no. 8, pp. 2783–2787, Feb. 2008, doi: 10.1073/pnas.0711990105.

[54] X. Huang *et al.*, "2D/3D heterojunction engineering at the grain boundaries towards high-performance inverted MA-free perovskite solar cells," *Org Electron*, vol. 122, p. 106918, Nov. 2023, doi: 10.1016/j.orgel.2023.106918.

[55] A. Amat *et al.*, "Cation-Induced Band-Gap Tuning in Organohalide Perovskites: Interplay of Spin–Orbit Coupling and Octahedra Tilting," *Nano Lett*, vol. 14, no. 6, pp. 3608–3616, Jun. 2014, doi: 10.1021/nl5012992.

[56] A. P. Grosvenor, M. C. Biesinger, R. S. C. Smart, and N. S. McIntyre, "New interpretations of XPS spectra of nickel metal and oxides," *Surf Sci*, vol. 600, no. 9, pp. 1771–1779, 2006, doi: 10.1016/j.susc.2006.01.041.

[57] M. C. Biesinger, B. P. Payne, A. P. Grosvenor, L. W. M. Lau, A. R. Gerson, and R. S. C. Smart, "Resolving surface chemical states in XPS analysis of first row transition metals, oxides and hydroxides: Cr, Mn, Fe, Co and Ni," *Appl Surf Sci*, vol. 257, no. 7, pp. 2717–2730, 2011, doi: 10.1016/j.apsusc.2010.10.051.

[58] A. J. Tkalych, K. Yu, and E. A. Carter, "Structural and Electronic Features of β-Ni(OH)2 and β-NiOOH from First Principles," *Journal of Physical Chemistry C*, vol. 119, no. 43, pp. 24315–24322, 2015, doi: 10.1021/acs.jpcc.5b08481.

[59] K. S. Kim and R. E. Davis, "Electron spectroscopy of the nickel-oxygen system," *J Electron Spectros Relat Phenomena*, vol. 1, no. 3, pp. 251–258, 1972, doi: 10.1016/0368-2048(72)85014-X.

[60] K. S. Kim, "SURFACE SCIENCE 43 (1974) 625-643 o North-Holland Publishing Co.," vol. 43, pp. 625–643, 1974.

[61] J. Landoulsi *et al.*, "Organic adlayer on inorganic materials: XPS analysis selectivity to cope with adventitious contamination," *Appl Surf Sci*, vol. 383, pp. 71–83, 2016, doi: 10.1016/j.apsusc.2016.04.147.

[62] D. Fang, F. He, J. Xie, and L. Xue, "Calibration of Binding Energy Positions with C1s for XPS Results," *Journal Wuhan University of Technology, Materials Science Edition*, vol. 35, no. 4, pp. 711–718, 2020, doi: 10.1007/s11595-020-2312-7.

[63] J. Tauc, "OPTICAL PROPERTIES AND ELECTRONIC STRUCTURE OF AMORPHOUS Ge AND Si," vol. 3, pp. 37–46, 1968.

[64] P. Gostishchev *et al.*, "Cl-Anion Engineering for Halide Perovskite Solar Cells and Modules with Enhanced Photostability," vol. 2200941, pp. 1–9, 2023, doi: 10.1002/solr.202200941.

[65] Y. Huang *et al.*, "The intrinsic properties of FA(1-:X)MAxPbI3 perovskite single crystals," *J Mater Chem A Mater*, vol. 5, no. 18, pp. 8537–8544, 2017, doi: 10.1039/c7ta01441d.

[66] C. C. Stoumpos, C. D. Malliakas, and M. G. Kanatzidis, "Semiconducting tin and lead iodide perovskites with organic cations: Phase transitions, high mobilities, and near-infrared photoluminescent properties," *Inorg Chem*, vol. 52, no. 15, pp. 9019–9038, 2013, doi: 10.1021/ic401215x.

[67] V. Campanari *et al.*, "Reevaluation of Photoluminescence Intensity as an Indicator of Efficiency in Perovskite Solar Cells," *Solar RRL*, vol. 6, no. 8, 2022, doi: 10.1002/solr.202200049.

[68]  V. Sarritzu *et al.*, "Optical determination of Shockley-Read-Hall and interface recombination currents in hybrid perovskites," *Nature Publishing Group*, no. February, pp. 1–10, 2017, doi: 10.1038/srep44629.

[69]  M. Stolterfoht *et al.*, "Visualization and suppression of interfacial recombination for high-efficiency large-area pin perovskite solar cells", doi: 10.1038/s41560-018-0219-8.

[70]  C. O. Martínez and R. L. Z. Hoye, "Chemie", doi: 10.1002/anie.202102360.

[71]  Y. Guo *et al.*, "Overcoming Ni $3+$ -Induced Non-Radiative Recombination at Perovskite-Nickel Oxide Interfaces to Boost Voltages in Perovskite Solar Cells," vol. 2100920, pp. 1–7, 2021, doi: 10.1002/admi.202100920.

[72]  H. Li *et al.*, "Buried Interface Dielectric Layer Engineering for Highly Efficient and Stable Inverted Perovskite Solar Cells and Modules," vol. 2300586, pp. 1–12, 2023, doi: 10.1002/advs.202300586.

[73]  C. C. Boyd, R. Cheacharoen, T. Leijtens, and M. D. McGehee, "Understanding Degradation Mechanisms and Improving Stability of Perovskite Photovoltaics," *Chem Rev*, vol. 119, no. 5, pp. 3418–3451, Mar. 2018, doi: 10.1021/ACS.CHEMREV.8B00336.

[74]  J. Xu *et al.*, "Triple-halide wide–band gap perovskites with suppressed phase segregation for efficient tandems," *Science (1979)*, vol. 367, no. 6482, pp. 1097–1104, Mar. 2020, doi: 10.1126/SCIENCE.AAZ5074.

[75]  C. C. Boyd *et al.*, "Overcoming Redox Reactions at Perovskite-Nickel Oxide Interfaces to Boost Voltages in Perovskite Solar Cells," *Joule*, vol. 4, no. 8, pp. 1759–1775, Aug. 2020, doi: 10.1016/J.JOULE.2020.06.004.

[76]  M. H. Futscher and C. Deibel, "Defect Spectroscopy in Halide Perovskites Is Dominated by Ionic Rather than Electronic Defects," *ACS Energy Lett*, vol. 7, no. 1, pp. 140–144, Jan. 2022, doi: 10.1021/acsenergylett.1c02076.

[77]  X. Ren *et al.*, "Deep-Level Transient Spectroscopy for Effective Passivator Selection in Perovskite Solar Cells to Attain High Efficiency over 23%," *ChemSusChem*, vol. 14, no. 15, pp. 3182–3189, Aug. 2021, doi: 10.1002/cssc.202100980.

[78]  S. Reichert, Q. An, Y.-W. Woo, A. Walsh, Y. Vaynzof, and C. Deibel, "Probing the ionic defect landscape in halide perovskite solar cells," *Nat Commun*, vol. 11, no. 1, p. 6098, Nov. 2020, doi: 10.1038/s41467-020-19769-8.

[79]  S. Heo *et al.*, "Deep level trapped defect analysis in $CH_3NH_3PbI_3$ perovskite solar cells by deep level transient spectroscopy," *Energy Environ Sci*, vol. 10, no. 5, pp. 1128–1133, 2017, doi: 10.1039/C7EE00303J.

[80]  N. Liu and C. Y. Yam, "First-principles study of intrinsic defects in formamidinium lead triiodide perovskite solar cell absorbers," *Physical Chemistry Chemical Physics*, vol. 20, no. 10, pp. 6800–6804, Mar. 2018, doi: 10.1039/C8CP00280K.

[81]  S. Tan, T. Huang, and Y. Yang, "Defect passivation of perovskites in high efficiency solar cells," *JPEn*, vol. 3, no. 4, p. 042003, Oct. 2021, doi: 10.1088/2515-7655/AC2E13.

[82]  M. Pols, J. M. Vicent-Luna, I. Filot, A. C. T. Van Duin, and S. Tao, "Atomistic Insights into the Degradation of Inorganic Halide Perovskite $CsPbI_3$: A Reactive Force Field Molecular Dynamics Study," *Journal of Physical Chemistry Letters*, vol. 12, no. 23, pp. 5519–5525, Jun. 2021, doi: 10.1021/ACS.JPCLETT.1C01192/SUPPL_FILE/JZ1C01192_SI_001.TXT.

[83]  K. Xue, C. Renaud, P. Y. Chen, S. H. Yang, and T. P. Nguyen, "Defect investigation in perovskite solar cells by the charge based deep level transient spectroscopy (Q-DLTS)," in *Lecture Notes in Networks and Systems*, vol. 63, Springer, 2019, pp. 204–209. doi: 10.1007/978-3-030-04792-4_28.

[84]  S. Reichert, J. Flemming, Q. An, Y. Vaynzof, J.-F. Pietschmann, and C. Deibel, "Improved evaluation of deep-level transient spectroscopy on perovskite solar cells reveals ionic defect distribution," Oct. 2019, Accessed: Mar. 09, 2020. [Online]. Available: http://arxiv.org/abs/1910.04583

[85] S. Srivastava *et al.*, "Advanced spectroscopic techniques for characterizing defects in perovskite solar cells," *Commun Mater*, vol. 4, no. 1, p. 52, Jul. 2023, doi: 10.1038/s43246-023-00379-y.

[86] C. Bi, X. Zheng, B. Chen, H. Wei, and J. Huang, "Spontaneous Passivation of Hybrid Perovskite by Sodium Ions from Glass Substrates: Mysterious Enhancement of Device Efficiency Revealed," *ACS Energy Lett*, vol. 2, no. 6, pp. 1400–1406, Jun. 2017, doi: 10.1021/acsenergylett.7b00356.

[87] X. Zheng *et al.*, "Dual Functions of Crystallization Control and Defect Passivation Enabled by Sulfonic Zwitterions for Stable and Efficient Perovskite Solar Cells," *Advanced Materials*, vol. 30, no. 52, Dec. 2018, doi: 10.1002/adma.201803428.

[88] Y. Li *et al.*, "Passivation of defects in perovskite solar cell: From a chemistry point of view," *Nano Energy*, vol. 77, p. 105237, Nov. 2020, doi: 10.1016/j.nanoen.2020.105237.

[89] Y. Yuan *et al.*, "Electric-Field-Driven Reversible Conversion Between Methylammonium Lead Triiodide Perovskites and Lead Iodide at Elevated Temperatures," *Adv Energy Mater*, vol. 6, no. 2, p. 1501803, Jan. 2016, doi: 10.1002/aenm.201501803.

[90] M. F. N. Taufique, R. Khanal, S. Choudhury, and S. Banerjee, "Impact of iodine antisite (IPb) defects on the electronic properties of the (110) CH3NH3PbI3 surface," *J Chem Phys*, vol. 149, no. 16, Oct. 2018, doi: 10.1063/1.5044667.

[91] S. G. Motti *et al.*, "Controlling competing photochemical reactions stabilizes perovskite solar cells," *Nature Photonics 2019 13:8*, vol. 13, no. 8, pp. 532–539, May 2019, doi: 10.1038/s41566-019-0435-1.

[92] G. Y. Kim, A. Senocrate, T.-Y. Yang, G. Gregori, M. Grätzel, and J. Maier, "Large tunable photoeffect on ion conduction in halide perovskites and implications for photodecomposition," *Nat Mater*, vol. 17, no. 5, pp. 445–449, May 2018, doi: 10.1038/s41563-018-0038-0.

[93] M. V. Khenkin *et al.*, "Consensus statement for stability assessment and reporting for perovskite photovoltaics based on ISOS procedures," *Nat Energy*, 2020, doi: 10.1038/s41560-019-0529-5.

[94] S. G. Motti *et al.*, "Defect Activity in Lead Halide Perovskites," *Advanced Materials*, vol. 31, no. 47, Nov. 2019, doi: 10.1002/ADMA.201901183.

[95] A. J. Barker *et al.*, "Defect-Assisted Photoinduced Halide Segregation in Mixed-Halide Perovskite Thin Films," *ACS Energy Lett*, vol. 2, no. 6, pp. 1416–1424, Jun. 2017, doi: 10.1021/acsenergylett.7b00282.

[96] S. G. Motti *et al.*, "Controlling competing photochemical reactions stabilizes perovskite solar cells," *Nat Photonics*, vol. 13, no. 8, pp. 532–539, Aug. 2019, doi: 10.1038/s41566-019-0435-1.

[97] C. Liu *et al.*, "Deep and shallow level defect passivation via fluoromethyl phosphonate for high performance air-processed perovskite solar cells," *Nano Energy*, vol. 118, p. 108990, Dec. 2023, doi: 10.1016/j.nanoen.2023.108990.

[98] F. Di Giacomo, L. A. Castriotta, F. U. Kosasih, D. Di Girolamo, C. Ducati, and A. Di Carlo, "Upscaling Inverted Perovskite Solar Cells: Optimization of Laser Scribing for Highly Efficient Mini-Modules," *Micromachines (Basel)*, vol. 11, no. 12, p. 1127, Dec. 2020, doi: 10.3390/mi11121127.

[99] N. H. Reich *et al.*, "Crystalline silicon cell performance at low light intensities," *Solar Energy Materials and Solar Cells*, vol. 93, no. 9, pp. 1471–1481, Sep. 2009, doi: 10.1016/j.solmat.2009.03.018.

[100] S. Kim, M. Jahandar, J. H. Jeong, and D. C. Lim, "Recent progress in solar cell technology for low-light indoor applications," *Current Alternative Energy*, vol. 03, Jan. 2019, doi: 10.2174/1570180816666190112141857.

[101] D. Saranin *et al.*, "Hysteresis-free perovskite solar cells with compact and nanoparticle NiO for indoor application," *Solar Energy Materials and Solar Cells*, 2021, doi: 10.1016/j.solmat.2021.111095.

[102] A. Sacco, L. Rolle, L. Scaltrito, E. Tresso, and C. F. Pirri, "Characterization of photovoltaic modules for low-power indoor application," *Appl Energy*, vol. 102, pp. 1295–1302, Feb. 2013, doi: 10.1016/J.APENERGY.2012.07.001.

[103] W. Tress, "Perovskite Solar Cells on the Way to Their Radiative Efficiency Limit – Insights Into a Success Story of High Open-Circuit Voltage and Low Recombination," *Adv Energy Mater*, 2017, doi: 10.1002/aenm.201602358.

[104] F. Wang, S. Bai, W. Tress, A. Hagfeldt, and F. Gao, "Defects engineering for high-performance perovskite solar cells," *npj Flexible Electronics*, 2018, doi: 10.1038/s41528-018-0035-z.

**Authors:**

P.K. Sukhorukova[1,2,§], E.A. Ilicheva[1,§], P.A. Gostishchev[1], L.O. Luchnikov[1], M.M. Tepliakova[3], D.O. Balakirev[2], I.V. Dyadishchev[2], A.A. Vasilev[4], D.S. Muratov[1], Yu. N. Luponosov[2], A. Di Carlo[5] and D.S. Saranin[1]

[1]LASE – Laboratory of Advanced Solar Energy, National University of Science and Technology "MISiS", Leninsky prospect 4, 119049 Moscow, Russia

[2] Enikolopov Institute of Synthetic Polymeric Materials of the Russian Academy of Sciences (ISPM RAS), Profsoyuznaya St. 70, Moscow, 117393, Russia

[3]Center for Energy Science and Technology, Skolkovo Institute of Science and Technology, Bolshoi blvd., 30, b. 1, Moscow, 121205, Russian Federation

[4]Department of Semiconductor Electronics and Semiconductor Physics, National University of Science & Technology MISIS, 4 Leninsky Ave., Moscow, 119049, Russia

[5]C.H.O.S.E. (Centre for Hybrid and Organic Solar Energy), Department of Electronic Engineering, University of Rome Tor Vergata, via del Politecnico 1, 00133 Rome, Italy

§: The authors contributed equally to this work

**Corresponding authors:**

Dr. Yu.N. Luponosov luponosov@ispm.ru, Dr. Danila S. Saranin saranin.ds@misis.ru, prof. Aldo Di Carlo aldo.dicarlo@uniroma2.it

# 1. Experimental section

## 1.1. Materials for synthesis of TPATC

4-Bromotriphenylamine, 2.5 M solution of *n*-butyllithium in hexane (*n*-BuLi) and tetrakis(triphenylphosphine)palladium (0) (Pd(PPh$_3$)$_4$) (Sigma-Aldrich Co) were used without further purification. Tetrahydrofuran (THF); toluene; ethanol and other solvents were purified and dried according to known methods. 4,4,5,5-Tetramethyl-2-(2-thienyl)-1,3,2-dioxaborolane (**2**) was prepared as described elsewhere [S1].

## 1.2. Materials for perovskite absorber and charge-transporting layers

Formamidinium iodide (FAI, >99.99%) was purchased from Greatcell Solar (Australia), Lead iodide (PbI$_2$, >99.9%) and cesium iodide (CsI, >99.99%) were purchased from Chemsynthesis (Russia) and LLC Lanhit (Russia), respectively. Fullerene-C$_{60}$ (C$_{60}$, >99.5%+) was purchased from MST (Russia). The organic solvents dimethylformamide (DMF), and isopropyl alcohol (IPA) were purchased in anhydrous from Sigma-Aldrich and used as received without further purification. Ethyl acetate (EAC) was purchased from HPS (Russia).

## 1.3. Synthetic procedures

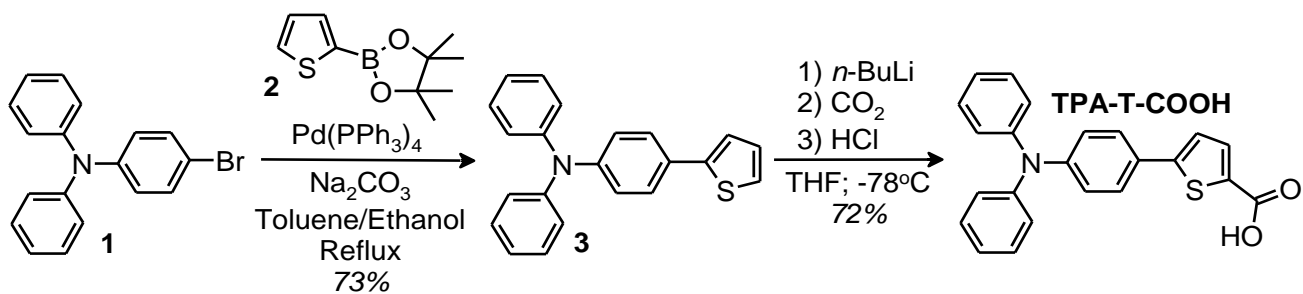

**Figure S1.** Synthesis of the **TPATC**.

**Diphenyl[4-(2-thienyl)phenyl]amine (3).** Degassed solutions of compound **1** (2.50 g, 7.71 mmol) and compound **2** (1.94 g, 9.25 mmol) in a toluene/ethanol mixture (60/10 mL) and 2 M aq. Na$_2$CO$_3$ solution (12 mL) were added to Pd(PPh$_3$)$_4$ (267 mg, 0.2 mmol). The reaction mixture was stirred under reflux for 6 h under an inert atmosphere. After completion of the reaction, 150 mL of toluene and 50 mL of distilled water were added to the reaction mixture. The organic phase was separated and washed with water. The solvent was evaporated under vacuum, and the crude product was purified by column chromatography (silica gel; eluent: toluene/hexane = 1/5) to give product **3** as a white powder (2.2 g, 73% yield). M.p. = 133 °C. $^1$H NMR (250 MHz, CDCl$_3$, δ, ppm): 7.01–7.16 (overlapping peaks, 9H); 7.20–7.32 (overlapping peaks, 6H); 7.47 (d, 2H, *J* = 8.66 Hz). $^{13}$C NMR (75 MHz, CDCl$_3$, δ, ppm): 122.21; 123.01; 123.75; 124.00; 124.41; 126.70; 127.97; 128.51; 129.29; 144.24; 147.17; 147.47. Calcd. (%) for C$_{22}$H$_{17}$NS: C, 80.70; H, 5.23; N, 4.28; S, 9.79. Found C, 80.96; H, 5.31; N, 4.12; S, 9.86. MALDI-TOF MS: found m/z 327.1; calcd. for [M]+ 327.1.

**5-[4-(Diphenylamino)phenyl]thiophene-2-carboxylic acid (TPATC).** *n*-BuLi (2.5 M solution in hexane, 1.1 mL, 2.7 mmol) was slowly added dropwise to a solution of compound **3** (0.9 g, 2.7 mmol) in 20 mL of dry THF at −78°C. After complete addition, the reaction mass was stirred at −78°C for 1 hour. Thereafter, excess dried CO$_2$ gas was slowly bubbled through the stirring reaction mass for 2 hours at −70°C. After the end of the reaction, the cooling bath was removed, and the stirring continued for another 30 minutes with the temperature rising to RT. After completion of the reaction, the mixture was poured in 100 mL of diethyl ether, containing 10 mL of 1 M HCl solution. The organic phase was washed with water (150 ml), then the organic phase was separated, and the solvent was evaporated under vacuum. After purification by recrystallization from ethanol, **TPATC** (0.74 g, 72%) was obtained as a light green powder. M.p. = 238 °C. $^1$H NMR (250 MHz, CDCl$_3$, δ, ppm): 6.93 (d, 2H, *J* = 8.80 Hz); 7.01–7.13 (overlapping peaks, 6H); 7.32 (t, 4H, *J* = 8.26 Hz); 7.41 (d, 1H, *J* = 3.85 Hz); 7.59 (d, 2H, *J* = 8.80 Hz); 7.66 (d, 2H, *J* = 3.85 Hz). $^{13}$C NMR (75 MHz, CDCl$_3$, δ, ppm): 122.32; 123.47; 123.88; 124.75; 126.32; 127.08; 129.79; 132.10; 134.54; 146.64; 147.96; 149.95; 162.94. IR

(cm$^{-1}$): 1660 (C=O). Calcd. (%) for C$_{23}$H$_{17}$NO$_2$S: C, 74.37; H, 4.61; N, 3.77; O 8.61; S, 8.63. Found C, 74.53; H, 5.24; N, 3.61; S, 8.69. MALDI-TOF MS: found m/z 370.9; calcd. for [M]+ 371.1.

### 1.4. Substrates

Solar cells were fabricated on In$_2$O$_3$:SnO$_2$ (ITO) coated glass (R$_{sheet}$<7 Ohm/sq) from Zhuhai Kaivo company (China). Optical measurements were done with use of KU-1 Quartz substrates from LLC Alkor (Russia). Thickness measurements were done with use of polished Si substrates from STV-Telecom (Russia).

### 1.5. Preparation of the Precursors

To fabricate the perovskite ink Cs$_{0.2}$FA$_{0.8}$PbI$_{3-x}$Cl$_x$ (text abbreviation CsFAPbI$_{3-x}$Cl$_x$) we used the halide salts (CsCl, CsI, FAI, PbI$_2$) with molar proportions 0.07:0.13:0.2:1. The resulting mixture was dissolved in a DMF:NMP(volume ratio 640:360) with a concentration of 1.35 M and stirred at a temperature of 50 °C for 1 h. BCP was dissolved in IPA at a concentration of 0.5 mg/ml and stirred for 12 h at 50 °C. Before use, BCP solution was filtered through 0.45 μm PTFE filters.

### 1.6. The Device Fabrication

The p-i-n-structured solar cell with the following stack ITO/NiO/CsFAPbI$_{3-x}$Cl$_x$/C60/BCP/Cu.

Firstly, the ITO substrates were cleaned with detergent, de-ionized water, acetone, and IPA in an ultrasonic bath. Then the substrates were activated under UV-ozone irradiation for 30 min. The NiO HTL was fabricated

The annealing of the deposited NiO layers was carried out on a heating plate in an air atmosphere with a relative humidity not exceeding 40%. CsFAPbI$_{3-x}$Cl$_x$ film was crystallized on top of HTL with a solvent-engineering method. The deposition and crystallization processes of perovskite layers were conducted inside glove boxes with an inert nitrogen atmosphere. The perovskite precursor was spin-coated with the following ramp: (1s – 1000 rpm, 4 sec – 3000 rpm / 30 sec – 5500 rpm). 420 μL of CB were dropped on the substrate on the 10$^{th}$ second after starting the first rotation step. Then the substrates were annealed at 70 °C (1 min) and 105°C (30 min) to form the appropriate perovskite phase. C$_{60}$ was deposited with the thermal evaporation method at 10$^{-6}$ Torr vacuum level. The free BCP interlayer was spin-coated at 4000 RPMs (30 s) and annealed at 50 °C (1 min) for the reference devices. The copper cathode was also deposited with thermal evaporation through a shadow mask to form a 0.14 cm$^2$ active area for each pixel.

We fabricated large-scale modules using laser beam patterning, employing a pulsed nanosecond laser (355 nm, 5 W from LLC NordLase, Russia) according to the approach presented in [S2].

### 1.7. Characterization

**NMR spectra.** $^1$H NMR spectra were recorded using a "Bruker WP-250 SY" spectrometer, working at a frequency of 250 MHz and using CDCl$_3$ (7.25 ppm) or DMSO-d6 (2.50 ppm) signals as the internal standard. $^{13}$C NMR spectra were recorded using a "Bruker Avance II 300" spectrometer, working at a frequency of 75 MHz. In the case of $^1$H NMR spectroscopy, the compounds to be analyzed were taken in the form of 1% solutions in CDCl$_3$ or DMSO-d6. In the case of $^{13}$C NMR spectroscopy, the compounds to be analyzed were taken in the form of 5% solutions in CDCl$_3$ or DMSO-d6. The spectra were then processed on the computer using the "ACD Labs" software.

**Elemental analysis.** Elemental analysis of C, N and H elements was carried out using CHN automatic analyzer "CE 1106" (Italy). The settling titration using BaCl$_2$ was applied to analyze the S element.

**Mass-spectra.** Mass-spectra (MALDI-TOF) were registered on a "Autoflex II Bruker" (resolution FWHM 18000), equipped with a nitrogen laser (work wavelength 337 nm) and time-of-flight mass-detector working in the reflections mode. The accelerating voltage was 20 kV. Samples were applied to a polished stainless-steel substrate. Spectrum was recorded in the positive ion mode. The resulting spectrum was the sum of 300 spectra obtained at different points of the sample. 2,5-Dihydroxybenzoic acid (DHB) (Acros, 99%) and *α*-cyano-4-hydroxycinnamic acid (HCCA) (Acros, 99%) were used as matrices.

**TGA.** Thermogravimetric analysis (TGA) was carried out in dynamic mode in 30 ÷ 600°C interval using a "Mettler Toledo TG50" system equipped with M3 microbalance allowing measuring the weight of samples in 0–150 mg range with 1 μg precision. Heating/cooling rate was chosen to be 10 °C/min. Every compound was studied twice: in the air and an under argon flow of 200 mL/min.

**DSC.** Differential scanning calorimetry (DCS) scans were obtained with a "Mettler Toledo DSC30" system with 20 °C/min heating/cooling rate in temperature range of +20–300 °C for all compounds. The $N_2$ flow of 50 mL/min was used.

**UV-Vis spectroscopy**. The absorption spectra were recorded with a "Shimadzu UV-2501PC" (Japan) spectrophotometer in the standard 10 mm photometric quartz cuvette using the $CHCl_3$ solution of the **TPATC** with the concentration of $10^{-5}$ M. All measurements were carried out at RT.

**CV.** Cyclic voltammetry (CV) measurements for **TPATC** film were carried out with a three-electrode electrochemical cell in an inert atmosphere in an electrolyte solution, containing 0.1 M tetrabutylammonium hexafluorophosphate ($Bu_4NPF_6$) in an acetonitrile and 1,2-dichlorobenzene (4:1) mixture using IPC-Pro M potentiostat. The scan rate was 200 mV s$^{-1}$. The glassy carbon electrode was used as the work electrode. The film was applied to a glassy carbon surface used as a working electrode by rubbing. A platinum plate placed in the cell served as the auxiliary electrode. Potentials were measured relative to a saturated calomel electrode (SCE). The highest occupied molecular orbital (HOMO) and the lowest unoccupied molecular orbital (LUMO) energy levels were calculated using the first formal oxidation and reduction potentials, respectively, obtained from CV experiments in acetonitrile according to following equations: LUMO = e($\varphi_{red}$+4.40) (eV) and HOMO = –e($\varphi_{ox}$+4.40) (eV) [S3, S4].

**Film characterization.** The film thickness and uniformity was measured with SENreserch 4.0 ellipsometer. The crystal structure of perovskite layers was investigated with X-ray diffractometer Difrey 401 using CrK1 as a source with wavelength 2.2904 Å under 20 kV voltage and a current of 8 mA. Wetting angle was measured using KRÜSS EasyDrop DSA20. The XPS measurements were performed using a «PREVAC EA15» electron spectrometer. In the current work, AlK$_\alpha$ (hv = 1486.6 eV, 150 W) were used as a primary radiation source. The pressure in analytical chamber did not exceed 5×10$^{-9}$ mbar during spectra acquisition. The binding energy (BE) scale was pre-calibrated using the positions of Ag3d$_{5/2}$ (368.3 eV) and Au4f$_{7/2}$ (84.0 eV) from silver and gold foils, respectively. We used The SmartSPM 1000 system (NSG30 tip) in AC mode for scanning atomic force microscopy characterization of NiO films. The optical properties were studied using a SE2030-010-DUVN spectrophotometer with a wavelength range of 200–1100 nm.

**JV-curves.** *JV*-curves were measured in an ambient atmosphere with Keithley 2401 SMU (voltage step of 23.5 mV and a settling time of 10$^{-2}$ s). The performance under 1 Sun illumination conditions were measured with ABET Sun3000 solar simulator (1.5 AM G conditions). Solar simulator was calibrated to standard conditions with a certified Si cell and an Ophir irradiance meter.

The dark *JV*-curves measurements were performed in the dark box.

**DLTS & AS.** Such measurements are particularly sensitive for device active layer properties and its effect of mobile ions. High concentration of mobile charged defects can screen build-in of applied fields on Debye length scale leading to device structure capacitance changes: $\left(C = \frac{\epsilon\epsilon_0 A}{L_D}, L_D = \sqrt{\frac{\epsilon\epsilon_0 k_B T}{q^2 N_i}}\right)$. The same approach used for determining times and frequencies for such defects will follow applied bias. For given diffusion coefficient $D$ time of mobile ion will travel across $L_D$ will be: $L_D^2/D = \tau$. So condition of peak in AS will be: $\omega \cdot \tau = 1 \Rightarrow \tau = \frac{\epsilon\epsilon_0 k_B T}{q^2 N_i D_0} \exp\left(\frac{E_A}{k_B T}\right)$, and then $E_A$ and $D_0$ can be determined from Arrhenius plot.

## 2. NMR spectra

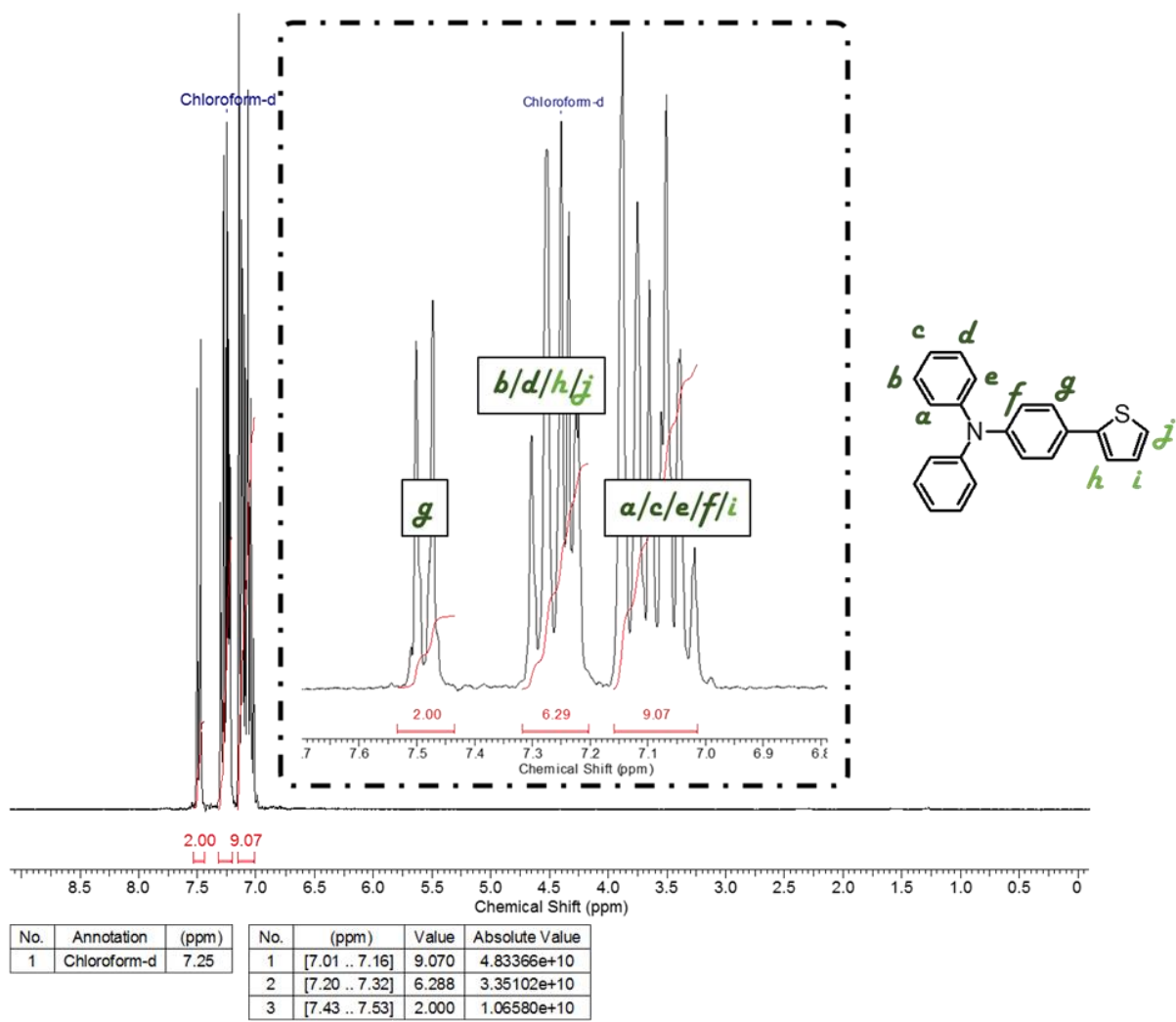

**Figure S2.** ¹H NMR spectrum of diphenyl[4-(2-thienyl)phenyl]amine (compound **3**)

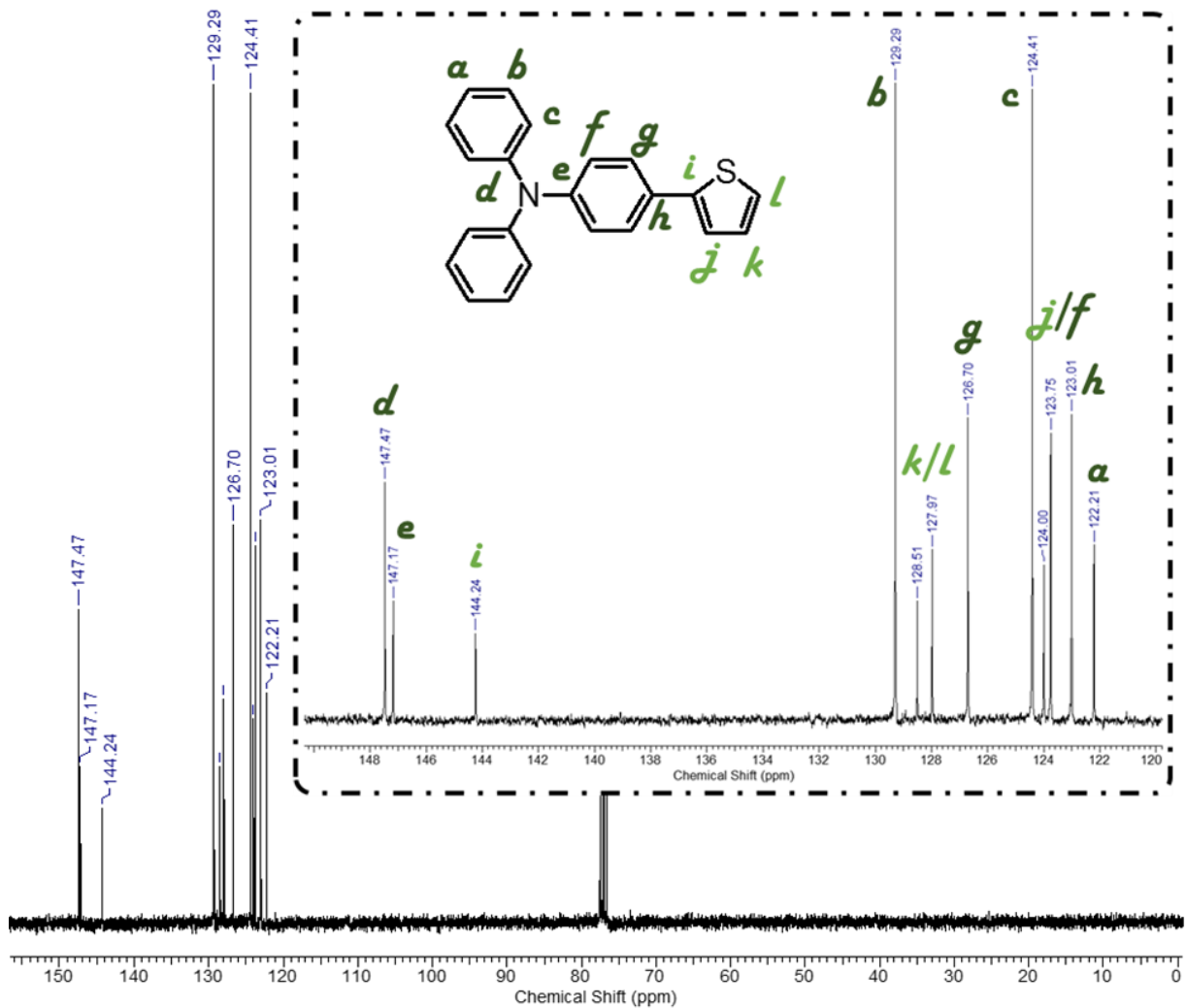

| No. | (ppm) | (Hz) | Height |
|---|---|---|---|
| 1 | 122.21 | 9224.1 | 0.2778 |
| 2 | 123.01 | 9284.6 | 0.4825 |
| 3 | 123.75 | 9340.5 | 0.4528 |
| 4 | 124.00 | 9359.0 | 0.2468 |
| 5 | 124.41 | 9390.2 | 0.9903 |
| 6 | 126.70 | 9563.3 | 0.4766 |
| 7 | 127.97 | 9659.2 | 0.2707 |
| 8 | 128.51 | 9699.3 | 0.1901 |
| 9 | 129.29 | 9758.2 | 1.0000 |
| 10 | 144.24 | 10886.6 | 0.1397 |
| 11 | 147.17 | 11108.3 | 0.1898 |
| 12 | 147.47 | 11130.4 | 0.3762 |

| No. | Annotation | (ppm) |
|---|---|---|
| 1 | Chloroform-d | [76.58 .. 77.43] |

**Figure S3.** $^{13}$C NMR spectrum of diphenyl[4-(2-thienyl)phenyl]amine (compound 3)

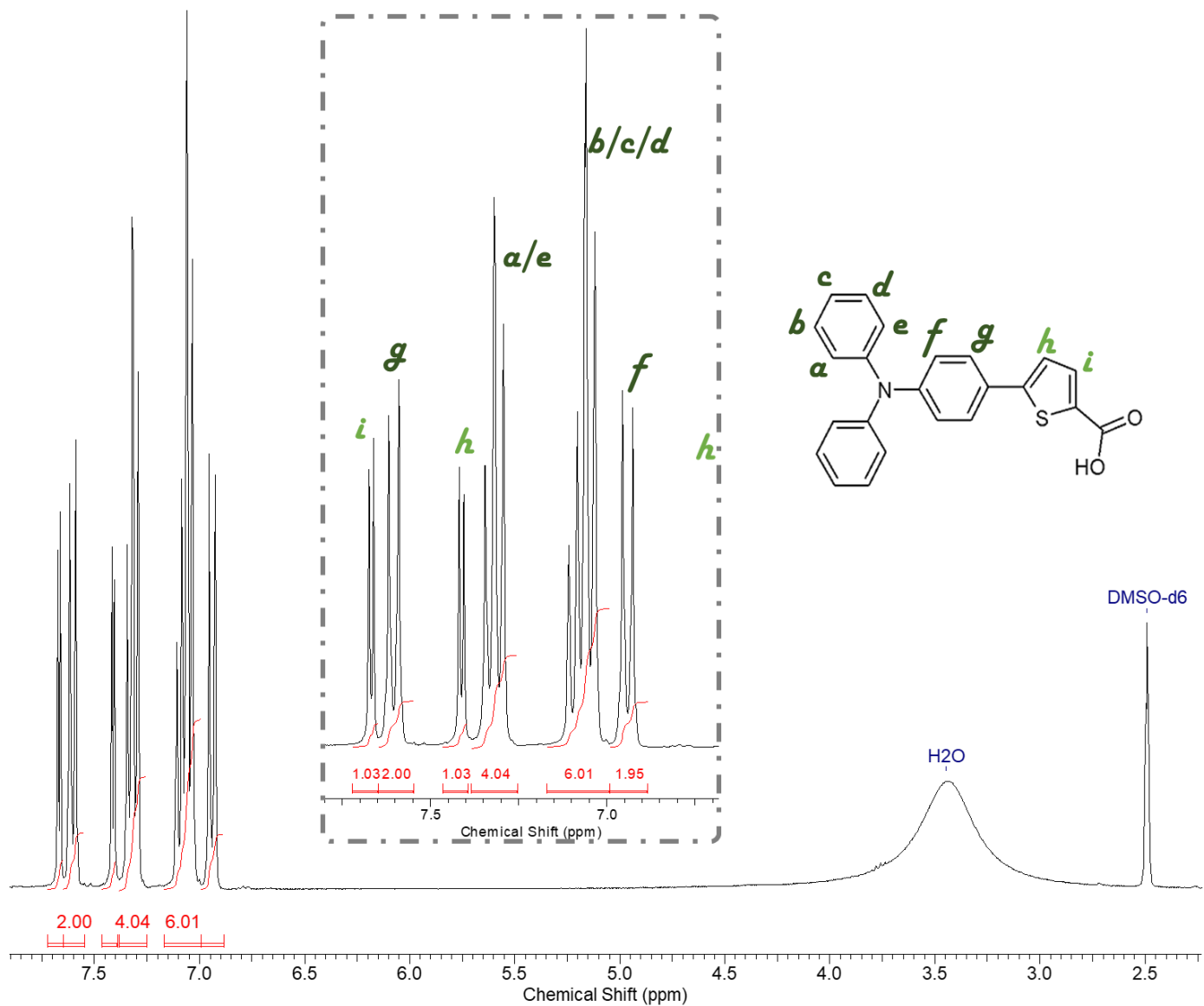

| No. | Annotation | (ppm) |
|---|---|---|
| 1 | DMSO-d6 | 2.49 |
| 2 | H2O | 3.45 |

| No. | (ppm) | Value | Absolute Value |
|---|---|---|---|
| 1 | [6.88 .. 6.99] | 1.949 | 5.47892e+9 |
| 2 | [6.99 .. 7.17] | 6.015 | 1.69071e+10 |
| 3 | [7.25 .. 7.38] | 4.040 | 1.13567e+10 |
| 4 | [7.39 .. 7.47] | 1.025 | 2.88155e+9 |
| 5 | [7.55 .. 7.65] | 2.000 | 5.62181e+9 |
| 6 | [7.65 .. 7.72] | 1.031 | 2.89678e+9 |

**Figure S4.** $^1$H NMR spectrum of **TPATC**

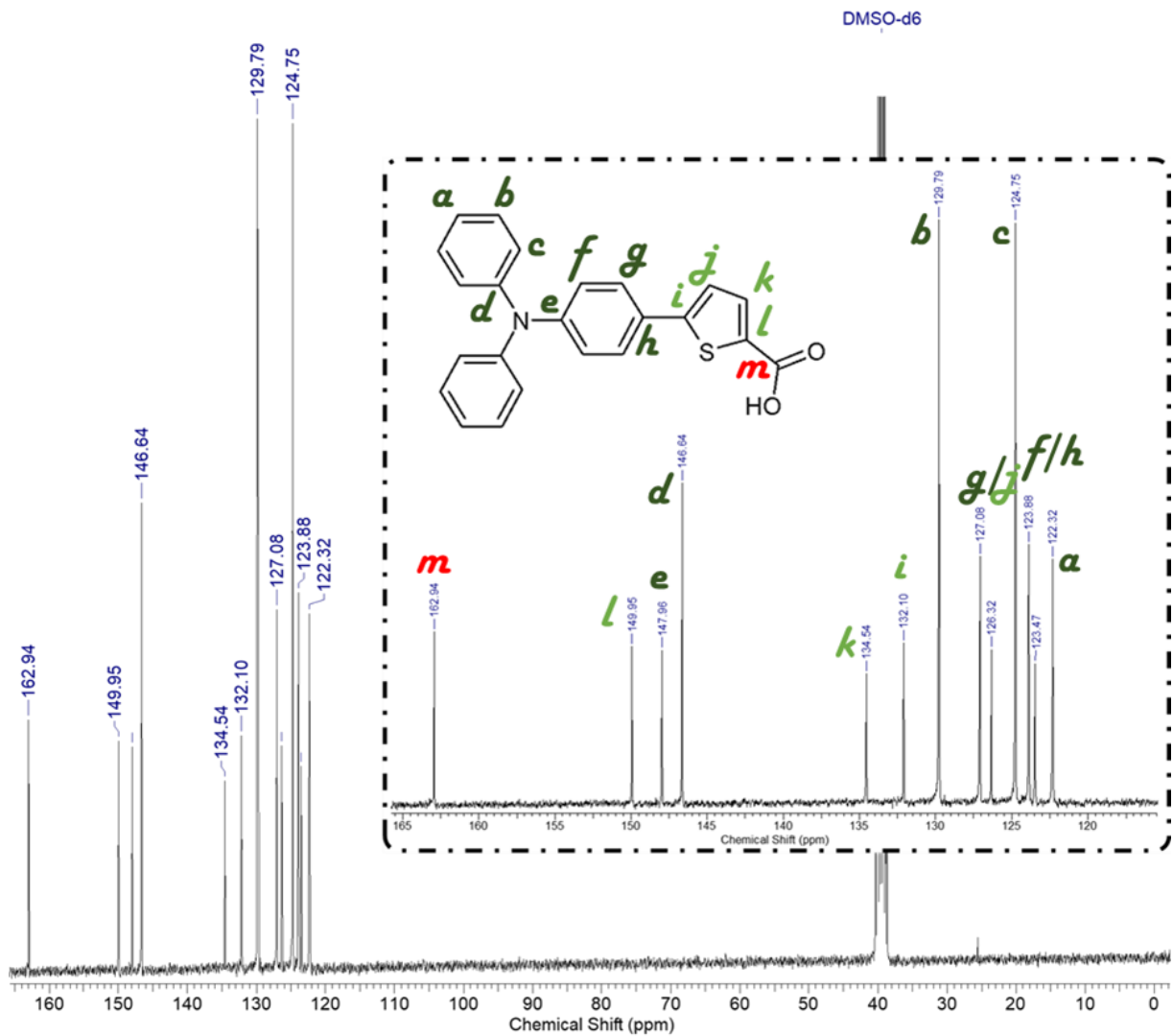

| No. | (ppm) | (Hz) | Height |
|---|---|---|---|
| 1 | 122.32 | 9232.1 | 0.1688 |
| 2 | 123.47 | 9318.9 | 0.0967 |
| 3 | 123.88 | 9350.5 | 0.1788 |
| 4 | 124.75 | 9415.7 | 0.4006 |
| 5 | 126.32 | 9534.6 | 0.1065 |
| 6 | 127.08 | 9591.5 | 0.1707 |
| 7 | 129.79 | 9796.1 | 0.4027 |
| 8 | 132.10 | 9970.8 | 0.1111 |
| 9 | 134.54 | 10154.9 | 0.0898 |
| 10 | 146.64 | 11068.2 | 0.2212 |
| 11 | 147.96 | 11167.7 | 0.1060 |
| 12 | 149.95 | 11318.1 | 0.1086 |
| 13 | 162.94 | 12298.4 | 0.1187 |

| No. | Annotation | (ppm) |
|---|---|---|
| 1 | DMSO-d6 | 39.50 |

**Figure S5.** $^{13}$C NMR spectrum of **TPATC**

## 3. FTIR spectrum

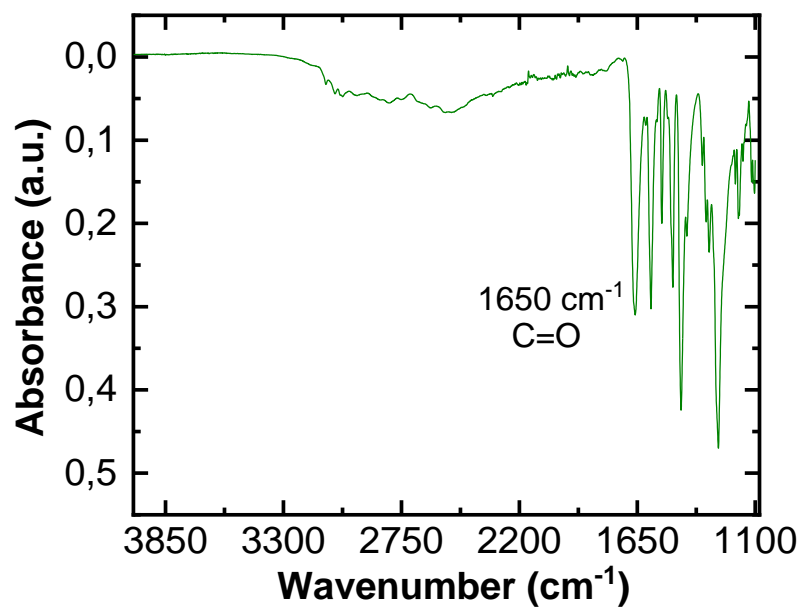

**Figure S6.** FTIR spectrum of **TPATC.**

## 4. Thermal properties and phase behavior

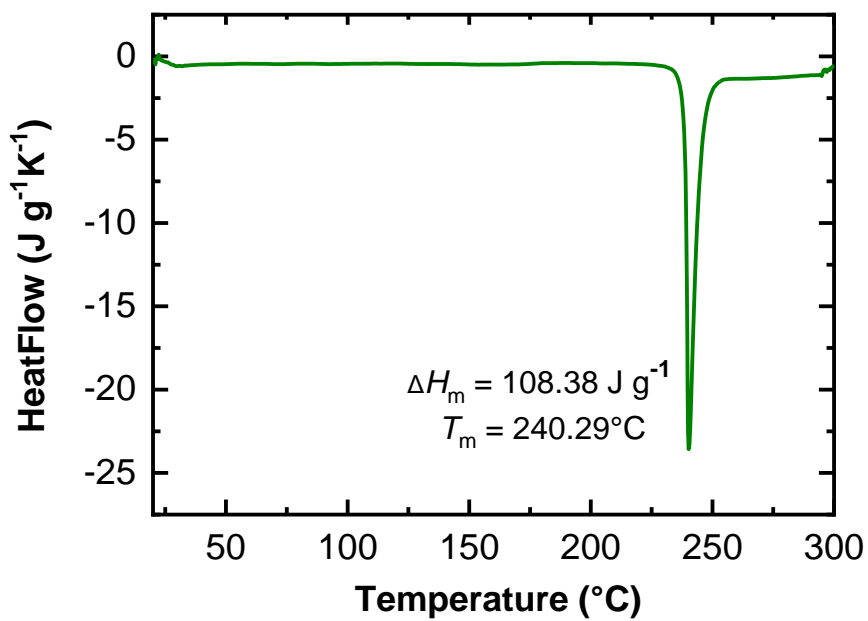

**Figure S7** DSC heating scan demonstrating the melting process for **TPATC.**

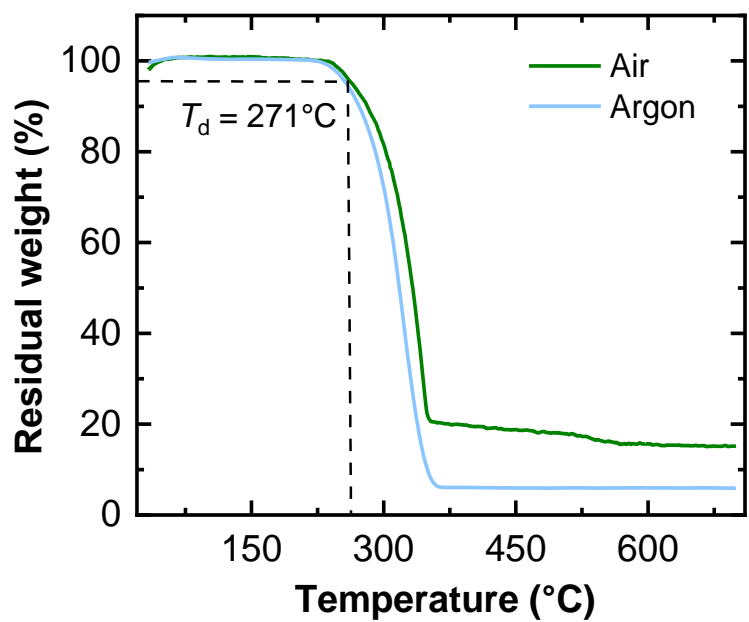

**Figure S8.** TGA curves of **TPATC** taken in an inert argon atmosphere and in the atmosphere.

5. Optical and electrochemical properties

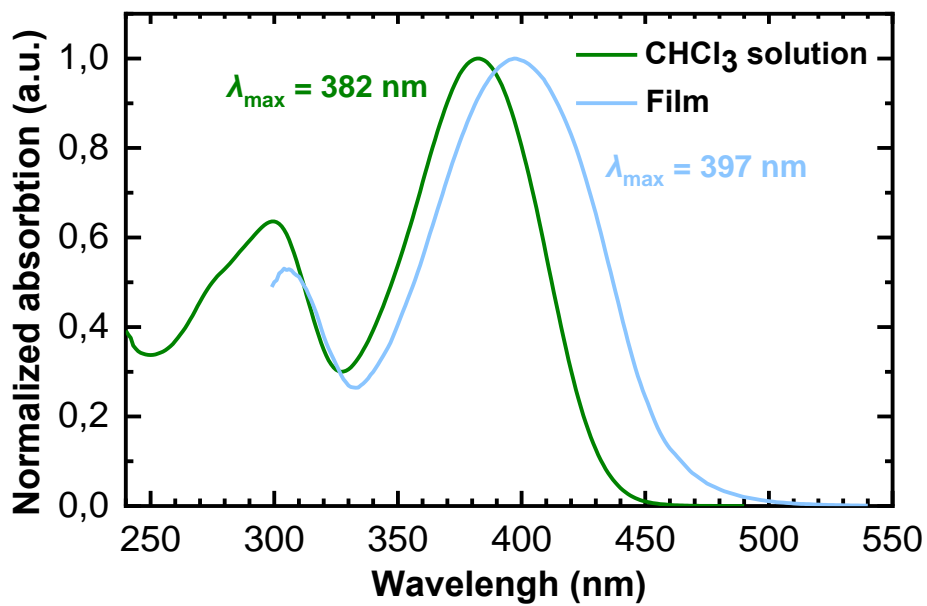

**Figure S9.** Normalized UV-vis absorption spectra of the **TPATC** in diluted chloroform solution and in thin film.

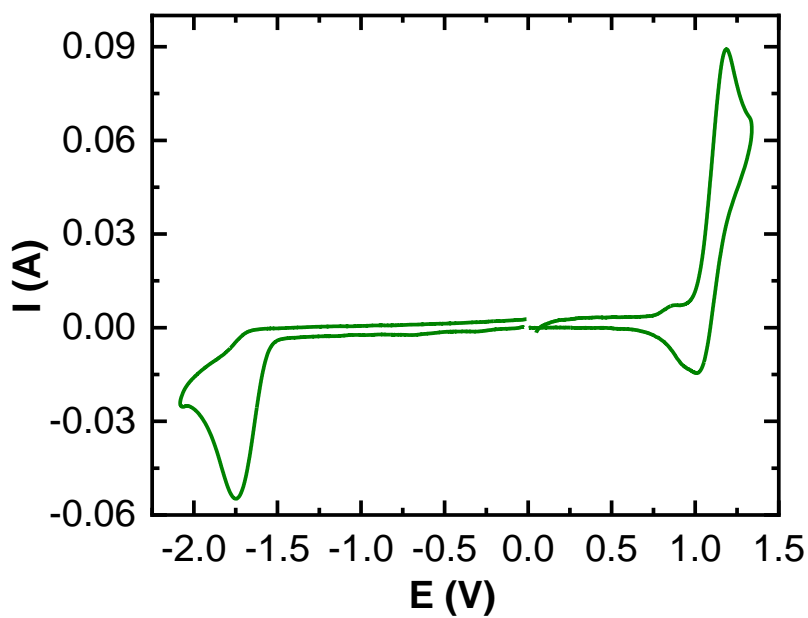

**Figure S10.** Cyclic voltammograms for thin polycrystalline film of **TPATC** in 1,2-dichlorobenzene/acetonitrile (1:4) mixture of solvents, which were recorded with a scan rate of 200 mV s$^{-1}$ using 0.1 M Bu$_4$NPF$_6$ as supporting electrolyte, glassy carbon (s = 2 mm$^2$) as work electrode, platinum plates as counter electrode and SCE (saturated calomel electrode) as reference electrode.

## 6. MALDI–TOF MS spectrum

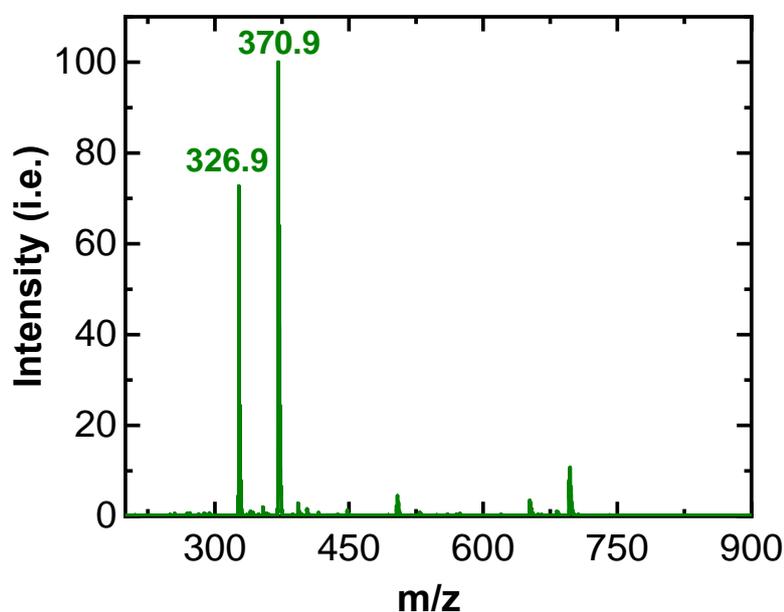

**Figure S11.** MALDI–TOF MS spectrum of **TPATC**. The spectrum contains two main signals with m/z values of 326.9 and 370.9, respectively. The second signal belongs to **TPATC**, while the first one corresponds to the theoretical one for diphenyl[4-(2-thienyl)phenyl]amine (compound **3**), and its presence can be explained by the process of acid decarboxylation at the moment of ionization.

## 7. Properties of hole-transporting thin-films

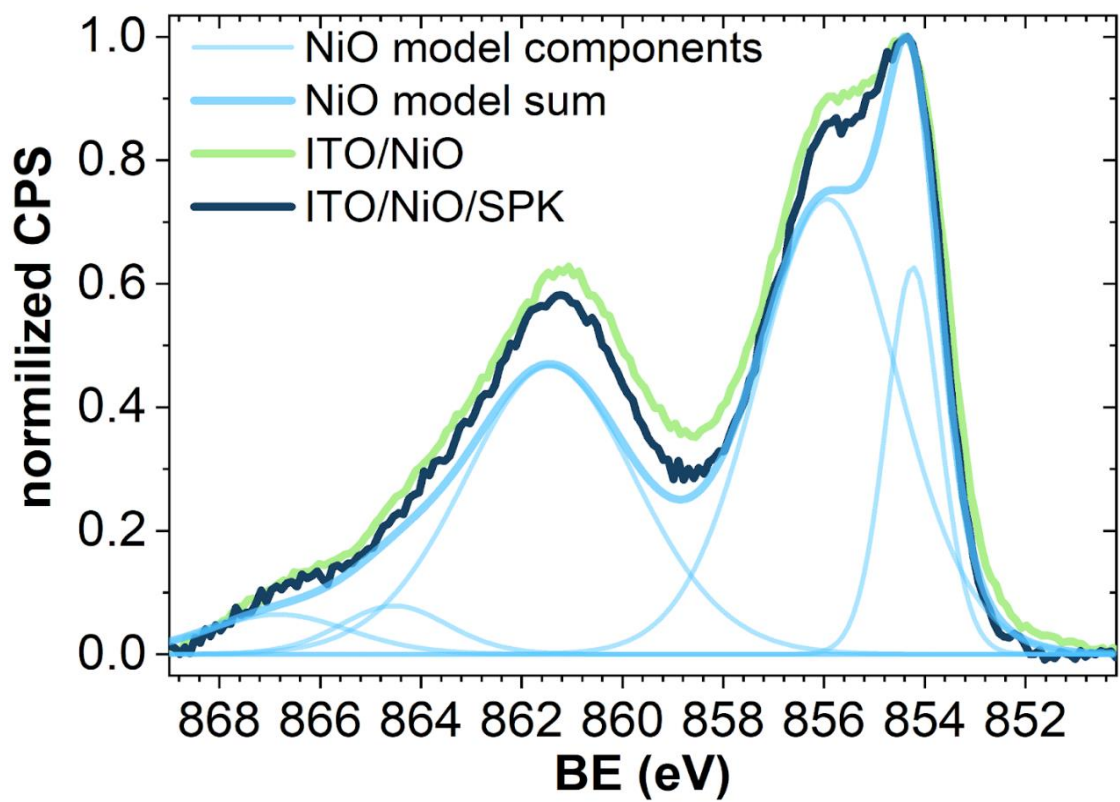

**Figure S12.** HR Ni2p core level of ITO/NiO and ITO/NiO/TPATC films

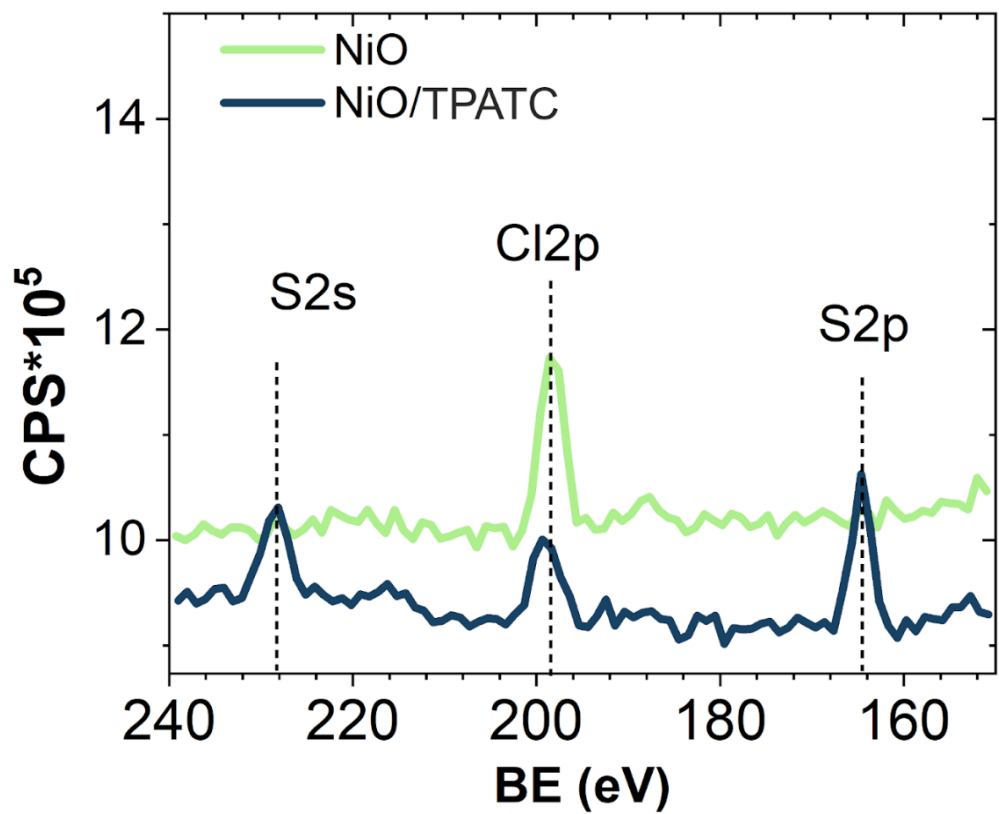

**Figure S13.** Survey photo emission spectra in Cl2p region for ITO/NiO and ITO/NiO/TPATC films

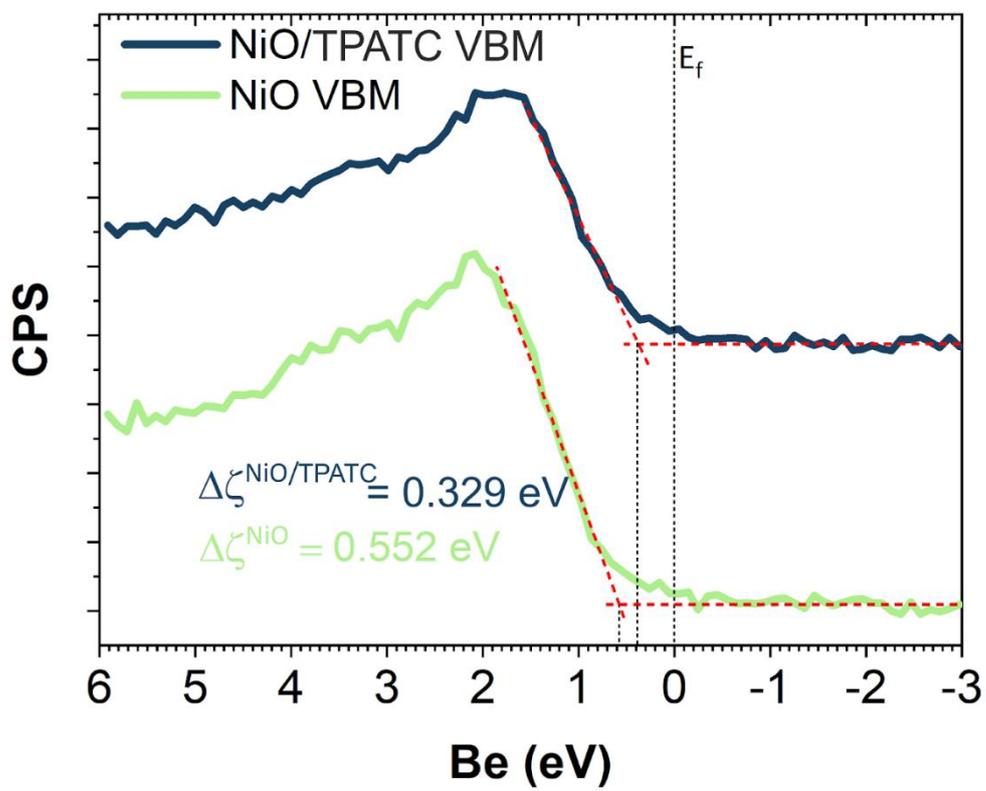

**Figure S15.** VBM spectra of ITO/ NiO and ITO/NiO/TPATC films near Fermi level

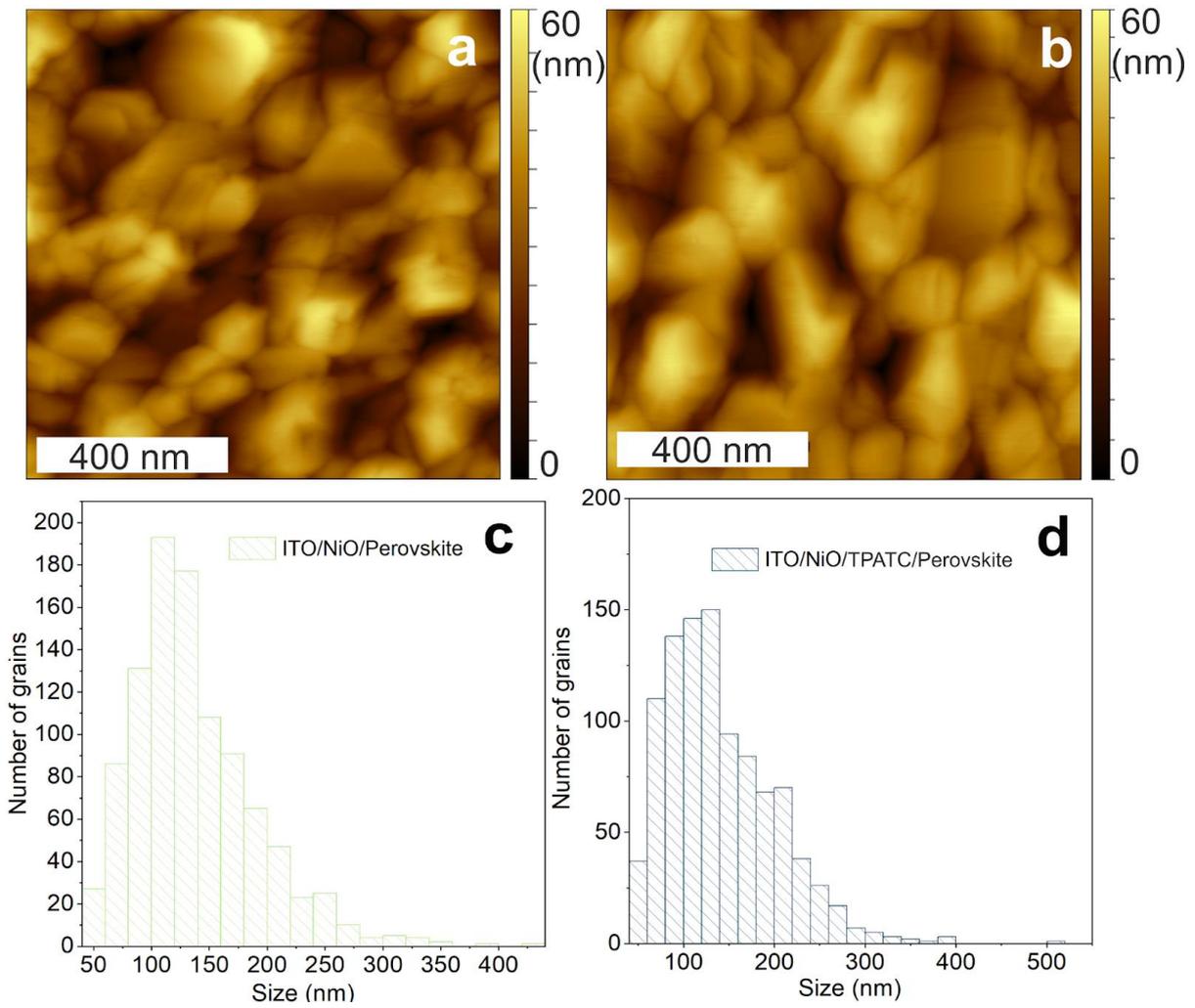

**Figure S16.** AFM of perovskite surfaces and grain size distribution for samples w/o TPATC (a, c) and with TPATC (b, d)

## 8. Device characteristics

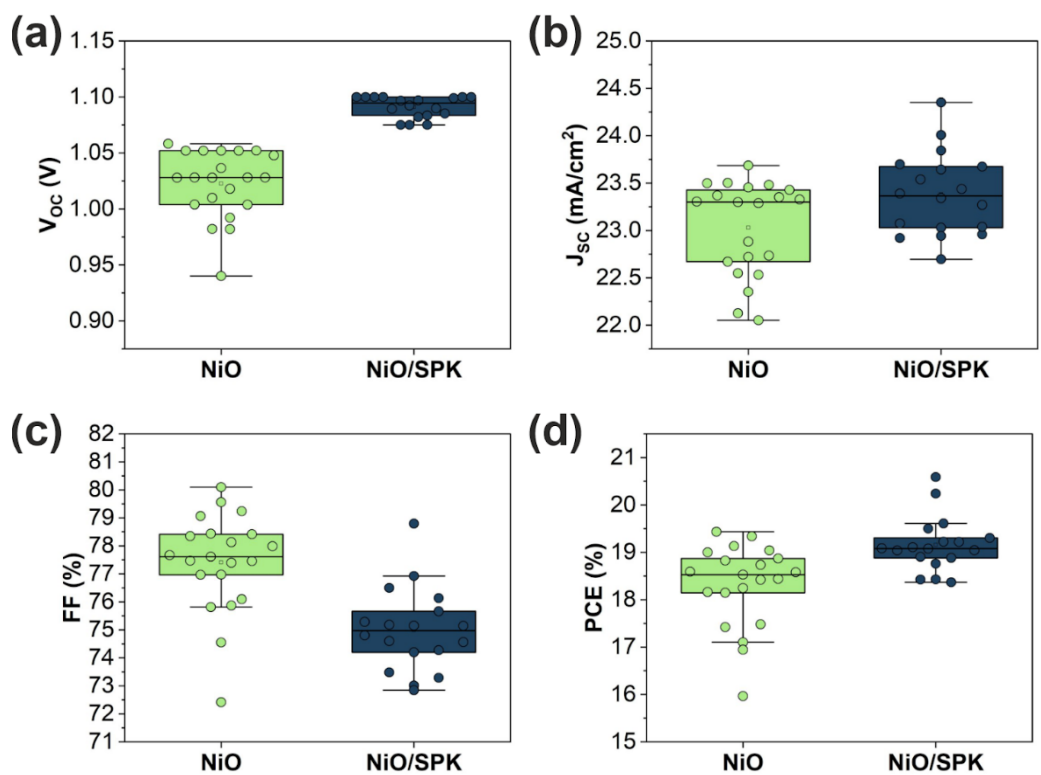

**Figure S17.** The box-charts with data extracted from the measured IV curves at standard conditions - $V_{OC}$ (a); $J_{SC}$ (b); FF (c) and PCE (d)

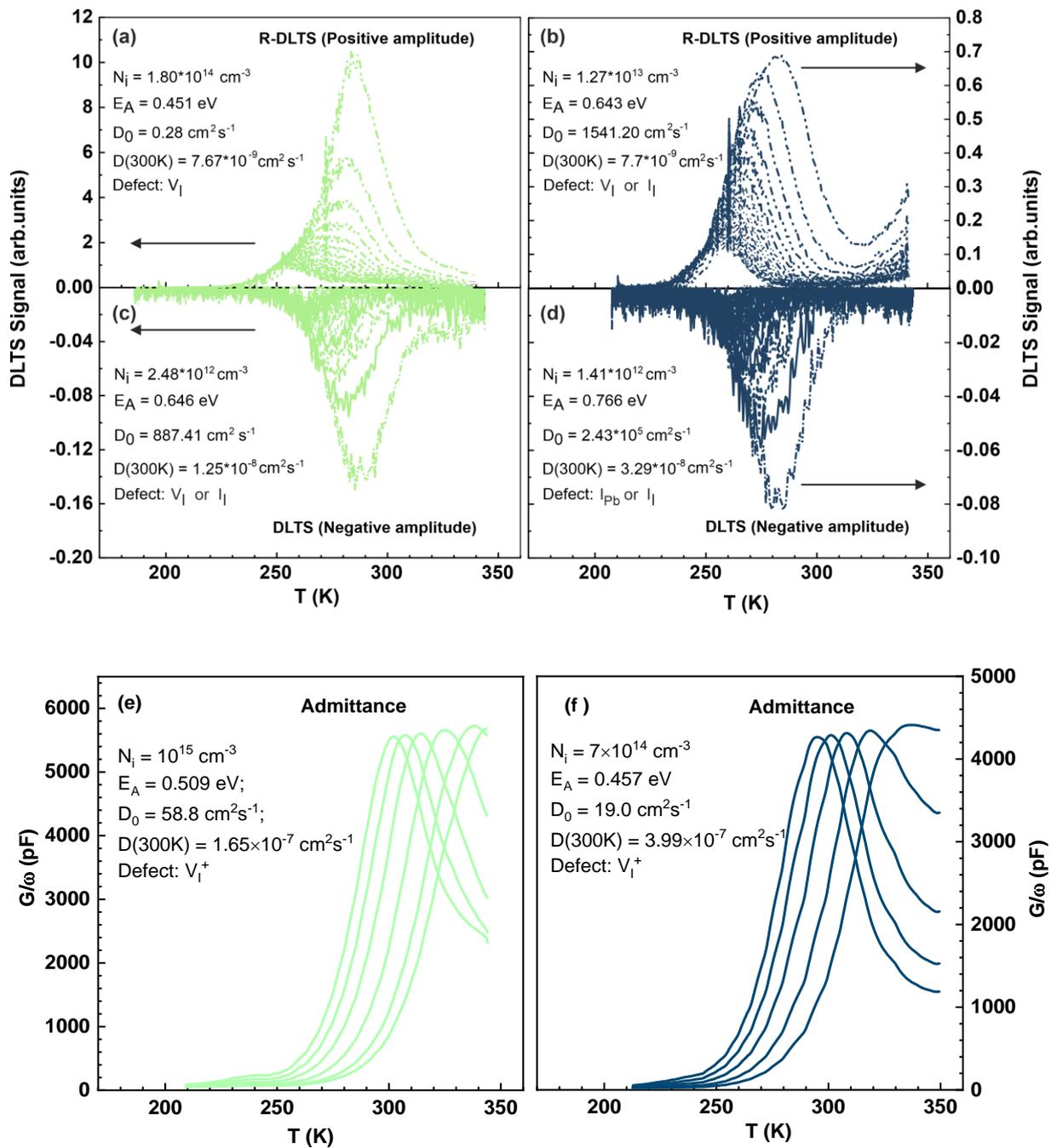

**Figure S18.** Deep level transient spectra for NiO PSCs with positive signal amplitudes (a), negative signal amplitudes (b), and for NiO/TPATC PSCs with positive signal amplitudes (c), negative signal amplitudes (d) and Admittance spectra for (e) NiO PSCs and (f) NiO/TPATC PSCs samples.

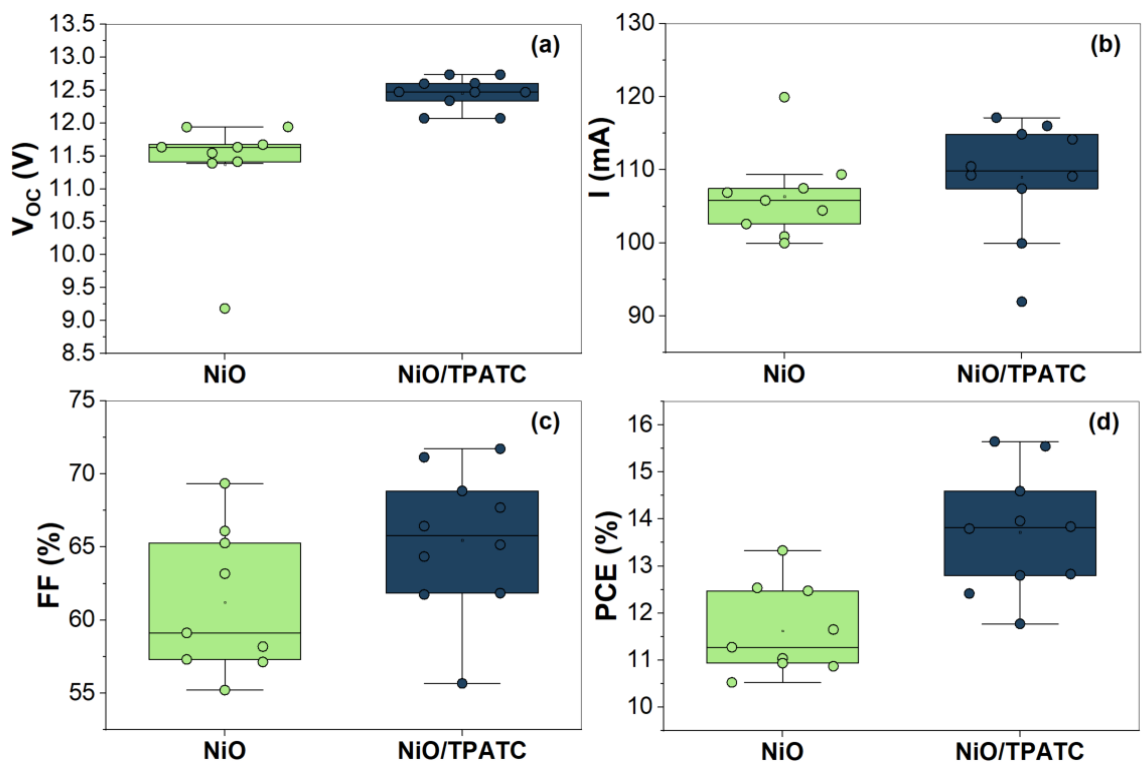

**Figure S19.** Output IV performance of the fabricated PSMs

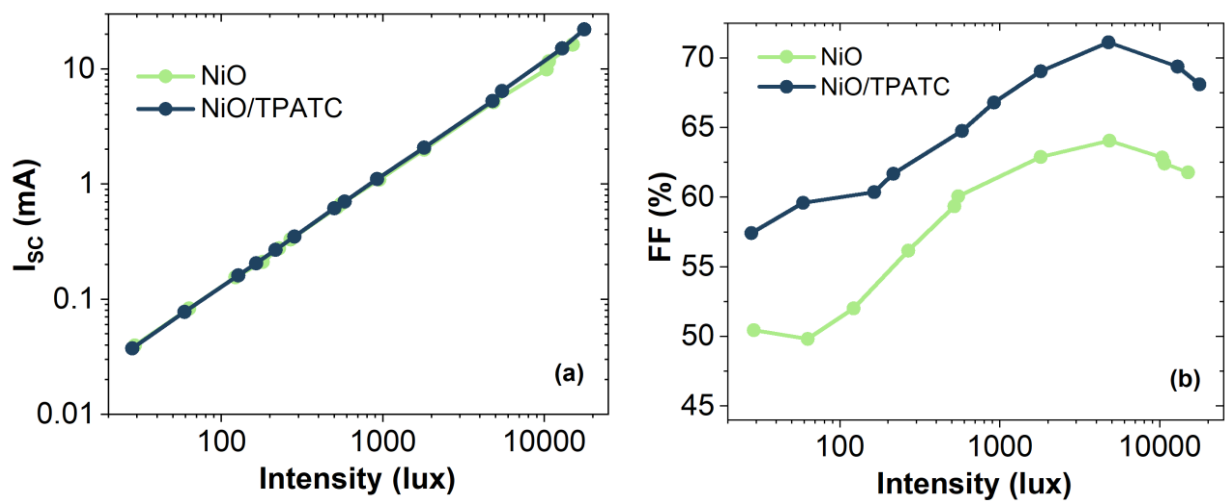

**Figure S20.** $I_{sc}$ and FF performance vs. $I_L$ under ambient low-light solutions

## 9. References

[S1] Y. Dienes, S. Durben, T. Kárpáti, T. Neumann, U. Englert, L. Nyulászi, T. Baumgartner, Selective Tuning of the Band Gap of π-Conjugated Dithieno[3,2-b:2′,3′-d]phospholes toward Different Emission Colors, Chemistry - A European Journal. 13 (2007) 7487–7500. https://doi.org/10.1002/chem.200700399.

[S2] C.M. Cardona, W. Li, A.E. Kaifer, D. Stockdale, G.C. Bazan, Electrochemical Considerations for Determining Absolute Frontier Orbital Energy Levels of Conjugated Polymers for Solar Cell Applications, Advanced Materials. 23 (2011) 2367–2371. https://doi.org/10.1002/adma.201004554.

[S3] S.A. Ponomarenko, N.N. Rasulova, Y.N. Luponosov, N.M. Surin, M.I. Buzin, I. Leshchiner, S.M. Peregudova, A.M. Muzafarov, Bithiophenesilane-Based Dendronized Polymers: Facile Synthesis and Properties of Novel Highly Branched Organosilicon Macromolecular Structures, Macromolecules. 45 (2012) 2014–2024. https://doi.org/10.1021/ma2024045.

[S4] F. Di Giacomo, L.A. Castriotta, F.U. Kosasih, D. Di Girolamo, C. Ducati, A. Di Carlo, Upscaling Inverted Perovskite Solar Cells: Optimization of Laser Scribing for Highly Efficient Mini-Modules, Micromachines. 11 (2020) 1127. https://doi.org/10.3390/mi11121127.